\documentstyle[aps,amstex,epsfig,harvard]{revtex}

\citationstyle{dcu}
\citationmode{abbr}

\newcommand{\illuseps}[3]{
\begin{figure}
\centerline{\epsfig{width=#1, file=#2.eps}}
\caption{#3 \label{#2}}
\end{figure}
}

\begin{document}

\title{The Effect of Lattice Vibrations on Substitutional Alloy Thermodynamics}
\author{A. van de Walle\\G. Ceder}
\address{Department of Materials Science and Engineering,\\
Massachusetts Institute of Technology, Cambridge, MA 02139}
\maketitle

\begin{abstract}
A longstanding limitation of first-principles calculations
of substitutional alloy phase diagrams is the difficulty to account for lattice vibrations.
A survey of the theoretical and experimental literature seeking to
quantify the impact of lattice vibrations on phase stability indicates that
this effect can be substantial. Typical vibrational entropy differences
between phases are of the order of $0.1$ to $0.2 k_B$/atom, which is comparable to
the typical values of configurational entropy differences in binary alloys
(at most $0.693 k_B$/atom).
This paper describes the basic formalism underlying {\it ab initio}
phase diagram calculations,
along with the generalization required to account for lattice vibrations.
We overview the various techniques allowing the theoretical calculation and
the experimental determination of phonon dispersion curves and related
thermodynamic quantities, such as vibrational entropy or free energy.
A clear picture of the origin of vibrational entropy differences between
phases in an alloy system is presented that goes beyond the traditional
bond counting and volume change arguments. Vibrational entropy change can be
attributed to the changes in chemical bond stiffness associated with the
changes in bond length that take place during a phase transformation. This
so-called ``bond stiffness vs. bond length'' interpretation both summarizes
the key phenomenon driving vibrational entropy changes and provides a
practical tool to model them.

Submitted to {\it Reviews of Modern Physics}.

\end{abstract}

\section{Introduction}

The field of first-principles alloy theory has made substantial progress over the last two
decades. It is now possible to predict relatively complex solid-state phase
diagrams starting from the basic principles of quantum mechanics and statistical mechanics.
Since no experimental input is required, these {\it ab-initio} calculations have been useful
for clarifying the phase diagram of several new materials \cite{ceder:ybacuo,vanderven:licoo2}.
Several excellent reviews on the topic exist
\cite{ducastelle:book,fontaine:clusapp,zunger:NATO}. The accuracy of calculated
phase diagrams is currently limited by two factors. First, one needs, as a
starting point, the energy of the alloy in various atomic configurations and
hence, one is limited by the accuracy of the quantum-mechanical calculations
used to obtain these energies. Typically, methods based on Density Functional Theory (DFT),
such as the Local
Density Approximation (LDA) or the Generalized Gradient Approximation (GGA) 
are used.

A second shortcoming arises from the fact that, in order to reduce computational requirements,
the sampling of the partition function to obtain the free energy is only done over
a limited number of degrees of freedom.
Typically, these include substitutional interchanges of atoms but no atomic vibrations.
Attempts to either assess the validity of this approximation or to devise
computationally efficient ways to account for lattice vibrations are
currently the focus of intense research. This interest is fueled by the
observation that phase diagrams obtained from first principles often
incorrectly predict transition temperatures. It is hoped that lattice
vibrations could account for some of the remaining discrepancies between theoretical
calculations and experimental measurements.

Three main questions are addressed in this paper.

\begin{enumerate}
\item  Do lattice vibrations have a sufficiently important impact on phase
stability that their thermodynamic effects need to be included in phase
diagram calculations?

\item  What are the fundamental mechanisms that explain the relationship
between the structure of a phase and its vibrational properties?

\item  How can the effect of lattice vibrations be modeled at a reasonable
computational cost?
\end{enumerate}

This paper is organized as follows. First, Section \ref{generalities}
presents the basic formalism that allows the calculation of phase diagrams,
along with the generalization needed to account for lattice vibrations. A
review of the theoretical and experimental literature seeking to
quantify the impact of lattice vibrations on phase stability is then
presented in Section \ref{evidence}. 
The main mechanisms through which lattice vibrations
influences phase stability are described in Section \ref{secmecha}.
Section \ref{compute}\ describes the
methods used to calculate vibrational properties while Section \ref
{experimental} presents the experimental techniques allowing their
measurement. Finally, Section \ref{modellatvib} discusses the strengths
and weaknesses of a variety of models of lattice vibrations.

\section{The formalism of alloy theory}

\label{generalities}Phase stability at constant temperature is determined by
the free energy\footnote{%
Strictly speaking, at constant pressure, the Gibbs free energy $G=F+PV$
should be used instead of the Helmoltz free energy $F$, but at atmospheric
pressure, the $PV$ term is negligible for an alloy.} $F$. The free energy
can be expressed as a sum of a configurational contribution $F_{\text{config}%
}$ and vibrational contributions $F_{\text{vib}}$. The configurational
contribution accounts for the fact that atoms can jump from one lattice site
to another, while vibrational contribution accounts for the vibrations of
each atom around its equilibrium position. The first part of this section
presents the traditional formalism used in alloy theory to determine the
configurational contribution. The second part introduces the basic
quantities that determine whether lattice vibrations have a significant
effect on phase stability. The third part describes how the traditional
formalism can be adapted when lattice vibrations do need to be accounted for.

\subsection{The cluster expansion formalism}

\label{alloyth}One of the goals of alloy theory is to determine the relative
stability of phases characterized by a distinct ordering of atomic species
on a given periodic array of sites. This array of sites, called the {\em %
parent lattice}, can be any crystallographic lattice augmented by any motif.
A convenient representation of an alloy system is the Ising model. In the
common case of a binary alloy system, the Ising model consists of assigning
a spin-like occupation variable $\sigma _{i}$ to each site $i$ of the parent
lattice, which takes the value $-1$ or $+1$ depending on the type of atom
occupying the site. A particular arrangement of spins of the parent lattice
is called a {\em configuration} and can be represented by a vector ${\bf%
\sigma}$ containing the value of the occupation variable for each site in
the parent lattice. Although this framework can be extended to arbitrary
multicomponent alloys \cite{sanchez:cexp}, we focus on the case of
binary alloys, since all the studies we review consider binary alloys only.

When all the fluctuations in energy are assumed to arise solely from
configurational change, the Ising model is a natural way to represent an
alloy. The thermodynamics of the system can then be summarized in a
partition function of the form: 

\begin{equation}
Z=\sum_{{\bf\sigma}}\exp \left( {-\beta E({\bf\sigma})}\right)
\label{pf_ising}
\end{equation}

where $\beta =1/\left( k_{B}T\right) $, and $E({\bf\sigma})$ is the energy
when the alloy has configuration ${\bf\sigma}$. It would be computationally
intractable to compute the energy of every configuration from
first-principles. Fortunately, the configurational dependence of the energy
can be parametrized in a compact form with the help of the so-called cluster
expansion \cite{sanchez:cexp}. The cluster expansion is a generalization of
the well-known Ising Hamiltonian. The energy (per atom) is represented as a
polynomial in the occupation variables: 
\begin{equation}
\frac{E({\bf\sigma})}{N}=\sum_{\alpha }m_{\alpha }J_{\alpha }\left\langle
\prod_{i\in \alpha ^{\prime }}\sigma _{i}\right\rangle  \label{clusexp}
\end{equation}
where $\alpha $ is a cluster (a set of sites $i$). The sum is taken over all
clusters $\alpha $ that are not equivalent by a symmetry operation of the
space group of the parent lattice, while the average is taken over all
clusters $\alpha ^{\prime }$ that are equivalent to $\alpha $ by symmetry.
The coefficients $J_{\alpha }$ in this expansion embody the information
regarding the energetics of the alloy and are called the effective cluster
interaction (ECI). The {\em multiplicities} $m_{\alpha }$ 
indicate the number of clusters that are equivalent by symmetry to $%
\alpha $ (divided by the number of lattice sites).\footnote{%
Both the number of clusters and the number of sites are infinite but their
finite ratio can be obtained by ignoring all but one periodic repetitions of
the clusters (or the atoms) by the a translational symmetry operation of the
lattice.}

It can be shown that when {\em all} clusters $\alpha $ are considered in the
sum, the cluster expansion is able to represent any function $E\left( {\bf%
\sigma}\right) $ of configuration ${\bf\sigma}$ by an appropriate selection
of the values of $J_{\alpha }$. However, the real advantage of the cluster
expansion is that, in practice, it is found to converge rapidly. A 
sufficient accuracy for phase diagram calculations can be achieved by keeping
only clusters $\alpha $ that are relatively compact ({\it e.g.} short-range
pairs or small triplets). The unknown parameters of the cluster expansion (the ECI) 
can then determined by fitting them to the energy of relatively small
number of configurations obtained, for instance, through
first-principles computations.
The cluster expansion thus presents an extremely
concise and practical way to model the configurational dependence of an
alloy's energy.

How many ECI and structures are needed in practice? A typical well-converged cluster
expansion of the energy of an alloy consists of about 20 to 30 ECI and necessitates the calculation
of the energy of around 40 to 50 ordered structures (see, for instance, \cite{vanderven:licoo2,gdg:linfit,ozolins:noble}).
A faithful modeling of the qualitative features of the phase diagram
(correct stable phases and topology) typically requires far fewer ECI (as little as 1 pair interaction) and
correspondingly less structures, as illustrated by the numerous examples given in \cite{fontaine:clusapp,ducastelle:book}.
In general multicomponent systems, the number of ECI and ordered structures
required to achieve a given precision 
unfortunately grows rapidly with the number of species (\cite{sanchez:cexp}).
For instance, in ternaries, each pair interaction is characterized by
3 interaction parameters instead of only one in the binary case.
For this reason, very few first-principle
calculations of ternary phase diagrams have been attempted
(see \cite{mccormack:cd2agau} for a recent example, or \cite{fontaine:clusapp} for a survey).

Although the cluster expansion usually allows a very compact representation of the energetics of an alloy system,
there are two situations where a standard cluster expansion is known to converge slowly.
Systems where long-range elastic interactions are important due to a large atomic size mismatch
between the alloyed species 
may require that elastic interactions be explicitly accounted for through the use of a so-called reciprocal space
cluster expansion \cite{zunger:NATO,ozolins:elas,ozolins:noble}.
Another situation, as recently identified by \cite{johnson:eta2}, is when the
electronic structure of the system exhibits a very strong configurational dependence due to
symmetry-breaking effects.

In the cases where a short-range cluster expansion does provide a sufficient accuracy,
the process of calculating the phase diagram of an alloy system can be
summarized as follows. First, the energy of the alloy in a relatively small
number of configurations is calculated, for instance through
first-principles computations. Second, the calculated energies are used to
fit the unknown coefficients of the cluster expansion (the ECI $J_{\alpha }$%
). Finally, with the help of this compact representation, the energy of a
large number of configurations is sampled, in order to determine the phase
boundaries. This latter step can be accomplished with either the Cluster Variation
Method (CVM) \cite{kikuchi:cvm,ducastelle:book}, the low-temperature
expansion (LTE) \cite{afk:lte}, or Monte-Carlo simulations \cite{binder:mc}.

\subsection{The effect of lattice vibrations}

The previous Section described the framework allowing the calculation of
phase diagrams under the assumption that the thermodynamics of the alloy is
determined solely by configurational excitations. Accounting for vibrational
excitations introduces corrections to this simplified treatment. This
section presents the basic quantities that enable an estimation of the
magnitude of the effect of lattice vibration on alloy thermodynamics. To
understand the effect of lattice vibrations on phase stability, it is
instructive to decompose the configurational (``config'') and vibrational
(``vib'') parts of the free energy $F^{\alpha }$ of a phase $\alpha $ into
an energetic contribution $E$ and an entropic contribution $S$: 
\begin{equation}
F^{\alpha }=E_{\text{config}}^{\alpha }-TS_{\text{config}}^{\alpha }+E_{%
\text{vib}}^{\alpha }-TS_{\text{vib}}^{\alpha }.  \label{EmTS}
\end{equation}
We take the convention that $E_{\text{config}}^\alpha$ is the energy of the alloy
system when all atoms are frozen at their average position at a given temperature.
In the approximation of harmonic lattice vibrations and in the limit of high
temperature, the vibrational energy $E_{\text{vib}}^{\alpha }$ is simply
determined by the equipartition theorem and is independent of the phase $%
\alpha $ considered. Hence as long as these approximations are appropriate,
lattice vibrations are mainly expected to influence phase stability through
their entropic contribution $S_{\text{vib}}^{\alpha }$.

Intuitively, the vibrational entropy $S_{\text{vib}}^{\alpha }$ is a measure
of the average stiffness of an alloy, as can be best illustrated by
considering an simple system made of large number of identical harmonic
oscillators. The softer the oscillators are, the larger their oscillation
amplitude can be, for a fixed average energy per oscillator. Hence, the
system samples a larger number of states and the entropy of the system
increases. In summary, the softer the alloy, the larger the vibrational
entropy.

A phase with a large vibrational entropy is stabilized relative to other
phases, since a larger vibrational entropy results in a lower free energy,
as seen by Equation (\ref{EmTS}). From a statistical mechanics point of
view, this fact can be understood by observing that a phase that encloses
more states in phase space is more likely to be visited, as the system
undergoes microscopic transitions, and therefore exhibits an increased
stability.

The central role of vibrational entropy can be further appreciated by
considering the effect of vibrations on a phase
transition between two phases $\alpha $ and $\beta $
which differ only by their average configuration
({\it e.g.} an order-disorder transition). 
If the vibrational entropy difference between the two phases is $\Delta S_{\text{vib}%
}^{\alpha \rightarrow \beta }$, the transition temperature obtained with
both configurational and vibrational contributions ($T_{\text{config+vib}%
}^{\alpha \rightarrow \beta }$) is related to the transition temperature
obtained with configurational effects only ($T_{\text{config}}^{\alpha
\rightarrow \beta }$) by 
\begin{equation}
T_{\text{config+vib}}^{\alpha \rightarrow \beta }\approx T_{\text{config}%
}^{\alpha \rightarrow \beta }\left( 1+\frac{\Delta S_{\text{vib}}^{\alpha
\rightarrow \beta }}{\Delta S_{\text{config}}^{\alpha \rightarrow \beta }}%
\right) ^{-1}  \label{tcshift}
\end{equation}
where $\Delta S_{\text{config}}^{\alpha \rightarrow \beta }$ is the change
in configurational entropy upon phase transformation \cite{gdg:vib1}.
This result is exact in the limit of small vibrational effects, high
temperature and harmonic vibrations. A correction to this result that
accounts for anharmonicity can be found in \cite{ozolins:cuau}.
Equation (\ref{tcshift}) indicates that the quantity
determining the magnitude of the effect of lattice vibration on phase
stability is the ratio of the vibrational entropy difference to the
configurational entropy difference. For this reason, most investigations
aimed at assessing the importance of lattice vibrations focus on estimating
vibrational entropy differences between phases. Since the configurational
entropy (per atom) $S_{\text{config}}$ for a binary alloy at concentration $c$ is bracketed by 
\begin{equation}
0\leq S_{\text{config}}\leq -k_{B}\left( c\ln c+\left( 1-c\right) \ln \left(
1-c\right) \right) \leq k_{B}\ln 2\approx 0.693k_{B},  \label{svibbnd}
\end{equation}
Equations (\ref{tcshift}) and (\ref
{svibbnd}) provide us with a absolute scale to gauge the importance of
vibrations. As we will see, typical vibrational entropy differences are of the
order of $0.2 k_B$, indicating that corrections of the
order of 30\% to the transition temperature may not be uncommon.

While it is clear that vibrational excitations introduce quantitative corrections to the simple picture
of alloy thermodynamics based on configurational excitations only, more profound
effects of a qualitative nature are also possible.
Vibrational effects may lead to deviations from the traditional
belief that, at high enough temperature, all short-range order in a disordered material
disappears \cite{gdg:vib1,Miller:vibrat}.
While a fully disordered state clearly maximizes configurational entropy $S_{\text{config}}$,
it is not clear that the total entropy $S_{\text{config}}+S_{\text{vib}}$ is necessarily
maximized in the state of maximum configurational disorder. The presence of short-range
order may {\em increase} the total entropy, relative to a fully disordered alloy,
through an increase of the vibrational entropy.
Vibrational entropy somewhat challenges our intuition, which is largely derived from
tacitly assuming that configurational disorder is {\em all} disorder.
It is even conceivable that vibrational entropy could
induce a transition from a disordered to an ordered phase with increasing temperature,
if the vibrational entropy difference between the ordered and disordered phases is larger
and opposite to the configurational entropy difference.
While this phenomenon has, so far, not been observed in metallic alloys, 
presumably because of the large configurational entropy associated
with disordering, it does occur in molecular systems, such as in diblock copolymer melts,
where the configurational entropy (per monomer) is small \cite{russell:lcot}.

\subsection{Coarse graining of the partition function}

\label{coarse}The cluster expansion formalism presented in Section \ref{alloyth} appears to
focus solely on configurational excitations. This section, shows that, in fact,
non-configurational sources of energy fluctuations can naturally be taken
into account within the cluster expansion framework through a process called
``coarse graining'' of the partition function\footnote{This ``coarse graining'' process is,
of course, unrelated to any processing aimed at increasing the grain size in a polycrystalline material.}
 \cite{ceder:phd,ceder:ising}.
This procedure also clarifies the nature of the physical states that are represented
by a configuration of the Ising model.

All the thermodynamic information of a system is contained in its partition
function: 
\begin{equation}
Z=\sum_{i}\exp \left[ -\beta E_{i}\right] ,  \label{pf_gen}
\end{equation}
where $E_{i}$ is the energy of the system in state $%
i $. In the case of a crystalline alloy system, the sum over all possible states of the
system can be conveniently factored as following: 
\begin{equation}
Z=\sum_{L}\sum_{{\bf\sigma}\in L}\sum_{v\in {\bf\sigma}}\sum_{e\in v}\exp %
\left[ -\beta E(L,{\bf\sigma},v,e)\right]  \label{factorize}
\end{equation}
where

\begin{itemize}
\item  $L$ is a so-called parent lattice: it is a set of sites where atoms
can sit. In principle, the sum would be taken over any Bravais lattice
augmented by any motif.

\item  ${\bf\sigma}$ is a configuration on the parent lattice: It specifies
which type of atom sits on each lattice site.

\item  $v$ denotes the displacement of each atom away from its ideal lattice
site.

\item  $e$ is a particular electronic state (both the spatial wavefunction
and spin state) when the nuclei are constrained to be in a state described
by $v$.

\item  $E(L,{\bf\sigma},v,e)$ is the energy of the alloy in a state
characterized by $L$, ${\bf\sigma}$, $v$ and $e$.
\end{itemize}

Each summation is taken over the states that are included in the set of
states defined by the ``coarser'' levels in the hierarchy of states. For
instance, the sum over displacements $v$ includes all displacements such
that the atoms remain close to the undistorted configuration ${\bf\sigma}$
on lattice $L$.

While Equation (\ref{factorize}) is in principle exact, practical
first-principles calculations of phase diagrams typically rely on various
simplifying assumptions. The sum over electronic states is often reduced to
a single term, namely, the electronic ground state. The validity of this
approximation can be assessed by ensuring that different structures
have a similar electronic density of states in the vicinity of the Fermi
level.\footnote{Unless strong electron correlation effects are present, such as charge ordering or
metal-insulator transitions \cite{imada:metalins},
the electronic free energy can be calculated from the single electron DOS
in the neighborhood of the Fermi level.}
If needed, the contribution of electronic entropy is, at least in its one-electron approximation,
relatively simple to include without prohibitive computational requirements 
\cite{wolverton:selec}.

A simplifying assumption that is much more difficult to relax is the reduction of
the sum over displacements $v$ to a single term. This simplification has
been extensively used in alloy theory, because calculating the summation
over $v$ involves intensive calculations. The particular displacement
representing a given configuration ${\bf\sigma}$ is typically chosen to be a
local minimum in energy that is close to the undistorted ideal
structure where atoms lie exactly at their ideal lattice sites. Usually, this state
is found by placing the atoms at their ideal lattice positions and relaxing the system
until a local minimum of the energy is obtained. In this
fashion, the state chosen is the most probable one in the neighborhood of
phase space associated with configuration ${\bf\sigma}$. In this
approximation, the partition function takes the form of an ising model
partition function: 
\begin{equation}
Z=\sum_{L}\sum_{{\bf\sigma}\in L}\exp (-\beta E^{\ast }(L,{\bf\sigma}))
\label{pf_ising2}
\end{equation}
with $E^{\ast }(L,{\bf\sigma})=\min_{v,e}\left\{ E(L,{\bf\sigma}%
,v,e)\right\} $.

It turns out that the same statistical mechanics techniques developed in the
context of the Ising model can also be used in the more general setting
where atoms are allowed to vibrate (and where electrons are allowed to be
excited above their ground state). All is needed is to replace the energy $%
E^{\ast }(L,{\bf\sigma})$ by the {\em constrained free energy} $F(L,{\bf%
\sigma},T)$, defined as: 
\begin{equation}
F(L,{\bf\sigma},T)=-k_{B}T\ln \left( \sum_{v\in {\bf\sigma}}\sum_{e\in
v}\exp \left[ -\beta E(L,{\bf\sigma},v,e)\right] \right) .  \label{consF}
\end{equation}
In other words, it is the free energy of the alloy, when its state in phase
space is constrained to remain in the neighborhood of the ideal
configuration ${\bf\sigma}$. This process, called the ``coarse graining''
of the partition function, is naturally interpreted as integrating out the
''fast'' degrees of freedom ({\it e.g.} vibrations) before considering
''slower'' ones ({\it e.g.} configurational changes) \cite{ceder:ising}.
This process is illustrated in Fig. \ref{coarsefig}.\ The quantity to be
represented by a cluster expansion is now the constrained free energy $F(L,%
{\bf\sigma},T)$. The only minor complication is that the effective cluster
interactions become temperature dependent.

There is some level of arbitrariness in the precise definition of the set of
displacement $v$ over which the summation is taken in Equation \ref{consF}.
However, in the common case where there is a local energy minimum in the
neighborhood of ${\bf\sigma}$ and where the system spends most of its time
visiting a neighborhood that can be approximated by a harmonic potential
well, the set of displacements over which the summation is taken has little
effect on the calculated thermodynamic properties. Under the above
assumptions, calculating the partition function of a constrained harmonic
system and a harmonic system that allows infinite displacements gives essentially the same
result: 
\begin{eqnarray*}
\sum_{v\in {\bf\sigma}}\exp \left[ -\beta E(L,{\bf\sigma},v,T)\right]
&\approx &\sum_{v\in {\bf\sigma}}\exp \left[ -\beta E_{H}(L,{\bf\sigma}%
,v,T)\right] \\
&\approx &\sum_{\text{all }v}\exp \left[ -\beta E_{H}(L,{\bf\sigma},v,T)%
\right]
\end{eqnarray*}
where $E(L,{\bf\sigma},v,T)=-k_{B}T\ln \sum_{e\in v}\exp \left[ -\beta E(L,%
{\bf\sigma},v,e)\right] $ and $E_{H}(L,{\bf\sigma},v,T)$ denotes a
harmonic approximation to $E(L,{\bf\sigma},v)$. In this framework, all that is
needed to account for lattice vibrations, is the determination of the free
energy of a harmonic solid in the neighborhood of any configurations ${\bf%
\sigma}$. (Appendix \ref{instab} discusses the case where the above
assumptions are violated, that is, when no local minimum exists in the phase
space neighborhood of ${\bf\sigma}$.)

The problem of calculating $F(L,{\bf\sigma},T)$ for a set of
configurations ${\bf\sigma}$ is much more demanding than
calculating the energy $E^{\ast }(L,{\bf\sigma})$ for a set of ${\bf%
\sigma}.$ Devising an efficient way to calculate $F(L,{\bf\sigma},T)$ is
the fundamental problem that needs to be resolved in order to include
vibrational effects in phase diagram calculations.

\subsection{Conclusion}

After presenting the basic framework enabling the calculation of the
configurational free energy,\ this section has presented two important
aspects of the thermodynamics of lattice vibrations.

\begin{enumerate}
\item  Vibrational entropy differences between phases introduce corrections
to transition temperatures calculated with only configurational entropy.

\item  The basic alloy theory framework can be adapted to account for
lattice vibrations by replacing the energy $E^{\ast }(L,{\bf\sigma})$
associated with each configuration ${\bf\sigma}$ by the free energy $F(L,%
{\bf\sigma},T)$ of a system constrained to remain in the phase space
neighborhood of the ideal configuration ${\bf\sigma}$.
\end{enumerate}

\section{Evidence of vibrational effects}

\label{evidence}In light of the large computational requirements associated
with the inclusion of lattice vibrations, is it important to ensure that
such an endeavor is worth the effort. This section reviews the experimental
and theoretical evidence that supports the view that vibrational effects are
important in the context of phase diagram calculations. There exists a large
literature aimed at determining the vibrational properties of solids (see,
for instance, \cite{ashmer:sol,maradudin:harm,born:phonon}). Here, the focus
is on investigations directly related to the determination of
vibrational entropy (or free energy) {\em differences} between phases which
differ solely by the ordering of the chemical species on an otherwise {\em %
identical parent lattice}.

This relatively narrow choice is driven by two observations. First, while
there have been numerous investigations of the absolute vibrational
properties of solids, the more difficult issue that needs to be addressed in
the context of phase stability is the determination of accurate {\em %
differences} in vibrational properties. Second, it has already been
established that many structural phase transformations ({\it e.g.}, from fcc
to bcc) are driven by lattice vibrations \cite{petry:bcc,grimvall:poly}.
This fact does not pose major difficulties for the purpose of phase diagram
calculations: one can easily compute the vibrational properties of a few
lattice types. The real difficulty is to calculate the vibrational entropy
of {\em many} configurations on {\em each} of these lattices, a task which
is only needed if vibrational properties differ substantially across
distinct configurations on an {\em identical parent lattice}.

The presentation will be mainly chronological, although deviations from that
intention will be made for the sake of clarity. We leave a more precise
description of the methods used for subsequent sections, focusing here on
the results obtained. The key theoretical and experimental results are
summarized in Tables \ref{tabsvibth} and \ref{tabsvibexp}, respectively.

\subsection{Calculations and predicted effects on phase stability}

\label{assesscalc}The idea that the state of order of an alloy could be
coupled with its lattice dynamics is not new. During the 1960's, as the
foundations of alloy thermodynamics were being established, the question of
the effect of lattice vibrations was already being raised. Studies on the
order-disorder transition of $\beta $-brass \cite{booth:latvib,wojto:latvib}%
, for instance, have indicated that lattice vibrations are crucial to
accurately model the magnitude of the experimentally observed discontinuity
in heat capacity at the phase transition, which determines the change in
vibrational entropy upon disordering. 

After these initial investigations, increasingly accurate models for the
coupling between lattice vibration and the state of order of an alloy, were
then developed \cite
{moraisis:tb,matthew:latvib,bakker:oddsvib,bakker:oddsvib2,bakker:isingvib,tuijn:odheat,gdg:vib1}%
. These models generally involved unknown parameters that need to be
estimated from available experimental thermodynamic data. A recurring theme
among these studies is the idea that, for sensible choices of the stiffness
of the springs connecting the atoms, the effect of lattice vibrations is
likely to be important. The estimated vibrational entropies of disordering
lie between $0.05k_{B}$, for the most conservative estimates \cite
{moraisis:tb}, up to the order of $0.5k_{B}$.\cite{tuijn:odheat}.

With the increased availability of computing power, the application of
first-principles methods became a practical possibility and the
unknown parameters of the theoretical models of lattice vibration have
become directly computable, without relying on experimental input.
Initially, only simple bulk properties, such as the bulk modulus were
computable at a reasonable cost. This prompted the development of methods to
infer vibrational properties from the knowledge of elastic constants. A
particularly popular scheme, the Moruzzi-Janak Schwarz (MJS) method \cite
{moruzzi:debyeg}, was used in many phase diagram calculations \cite
{asta:cdmg,sanchez:agcu,tseng:tial,sanchez:zrnb,Mohri:CoPt,Colinet:AuNi,Abrikosov:NiAl,Mohri:AuCu-AuPd-AuAg,Mohri:GaAs-InAs,Mohri:InP-InSb,Sanchez:RuZrNb}%
. In the Cd-Mg \cite{asta:cdmg}, Ag-Cu \cite{sanchez:agcu} and Au-Ni \cite
{Colinet:AuNi} systems, agreement with the experimentally measured phase
diagrams was substantially improved by including vibrations in this way.
In retrospect, such an improvement was to be expected because first-principles
phase diagram calculations often greatly overestimate transitions temperature
and {\em any} downward
correction to the calculated transition temperatures yields improved
agreement. The MJS approximation nearly always yields a downward correction
when ordered compounds are stiffer (softer) than the elements in ordering
(phase separating) systems, a very likely situation.

More recently, other techniques were used to obtain vibrational properties
from elastic constants. The so-called virtual crystal approximation (described
in Appendix \ref{appdisord})
was used to calculate the vibrational free energy of a disordered alloy in
the Ni-Cr system \cite{craievich:nicr}. The calculated vibrational free
energy exhibited qualitatively the same concentration-dependence as the
vibrational free energy obtained by subtracting the experimentally
determined free energy from the calculated configurational free energy.
The virtual crystal approximations was also used in calculations of
the lattice dynamics of ordered and disordered Cu$_{3}$Au  \cite{cleri:cu3au}.
These calculations, which relied on a tight-binding Hamiltonian (see, for instance,
\cite{Harrison:book,pettifor:tb}), predicted a $0.12k_{B}$ increase
in the vibrational entropy upon disordering.

In view of the large computational requirements of accurate {\it ab initio}
methods, many researchers have sought to calculate vibrational properties with
simpler energy models, whose lower computational requirements enable a more
accurate handling of issues such as anharmonicity or the representation of
the disordered state. The development of the embedded atom method (EAM) \cite
{daw:eam} offered the opportunity to accurately model metallic alloys at a
reasonable computational cost. Investigations based on the EAM have typically
found a large vibrational entropy change upon disordering in metallic alloys.
The vibrational entropy change for Cu$_{3}$Au was predicted to be $0.10k_{B}$
\cite{ackland:vol}, while for Ni$_{3}$Al, values ranging from $0.22k_{B}$ to 
$0.29k_{B}$ were obtained \cite
{ackland:vol,althoff:vib,althoff:vibcms,ravelo:vib}\footnote{%
Disordered Ni$_{3}$Al is actually a metastable phase. The values quoted here
are at the highest temperatures reported by the investigators, as close as
possible to the true disordering temperature that would be observed if the
alloy did not melt before.}.

Other researchers constructed pair potentials from an equation of state
determined from first-principles calculations \cite{shaojun:feal}. The
resulting pair potentials were then used to calculate disordering
vibrational entropies with no further approximations. This method attributed
a vibrational entropy change of $0.11k_{B}$ to the disordering reaction of
the Fe$_{3}$Al compound.

Rather unexpected results were uncovered as it became possible to compute
vibrational entropy differences from a complete lattice dynamics analysis as
well as state-of-the-art {\it ab initio} techniques. Calculations on the
Si-Ge system \cite{gdg:phd} found almost no effect of lattice vibrations:
the vibrational entropy of formation of the metastable zincblende structure
was a mere $-0.02k_{B}$. The first {\it ab initio} calculation of a
vibrational entropy of disordering \cite{avdw:ni3al} placed an upper bound
of $0.05k_{B}$ in the case of the order-disorder transition of the Ni$_{3}$%
Al compound, in sharp contrast with previous EAM calculations
\cite{ackland:vol,althoff:vib,althoff:vibcms} which found a
much larger value. Although the disagreement simply originated
from a difference in the predicted volume expansion of the alloy
upon disordering, the difference between EAM and {\it ab initio}
results does indicate that vibrational entropy differences
are very sensitive to the energy model used.
In another intermetallic compound, Pd$_{3}$V, the vibrational entropy
change upon disordering of the  \cite{avdw:pd3v}
was calculated to be $-0.07k_{B}$, although the simplest theories put
forward in the earliest investigations of lattice vibrations in alloys would
predict this value to be large and positive.

In the first phase diagram
calculation based on a full lattice dynamics analysis \cite{tepesch:caomgo},
the SCPIB method \cite{boyer:PIB1} was used to calculate that vibrational
effects lower the top of the miscibility gap by about $10$\% in the CaO-MgO
system.
Subsequently, first-principles calculations on the Cu-Au system \cite{ozolins:cuau}
reported a reduction of a about $20$\% in the transition temperatures.
However, in both cases, the resulting correction on the phase boundaries decreased the agreement with
the experimentally determined phase diagram, suggesting that other potential sources
of error, such as the precision of the energy method used or the
convergence of the cluster expansion, have to be investigated.

Very recently, two rather striking examples of systems where lattice vibrations
do have an important effect and where their inclusion results in a 
a dramatically improved agreement with experimental measurements have been
uncovered. Lattice vibrations are responsible for a 27-fold increase
in the solubility of scandium in aluminum \cite{ozolins:alsc}, resulting 
in a nearly perfect agreement with experimentally determined solubility limits.
Vibrational contributions were also shown \cite{wolverton:cual}
to be essential to correctly predict the relative stability of the different
precipitates which constitute the well-known Guinier-Preston zones responsible
for precipitation hardening in the Al-Cu system.
It is interesting to note that these two successful examples
did not require the use of a cluster expansion and did not necessitate a large number
of separate first-principles calculations. Each calculation could therefore be carried out
with an extremely high precision. When all other sources of errors are well controlled,
the effect of lattice vibrations can be more accurately quantified.
These examples offer good hope that {\it ab-initio} calculations of a complete phase diagram
that include lattice vibrations can provide an accuracy comparable to experiments, provided that
a well converged cluster expansion can be obtained and that highly accurate first-principles
calculations are used to construct it.

\subsection{Comparison with Experiments}

Over the last 10 years, advances in experimental techniques have made it
possible to directly measure vibrational entropy differences, instead of
inferring them from discrepancies between measured thermodynamical data and
calculated estimates of configurational contributions to the free energy.
Experimental investigations have thus provided independent assessments of
the role of lattice vibration on phase stability.

The first direct measurement was obtained from differential calorimetry
measurements on the Ni$_{3}$Al compound \cite{fultz:ni3alcal} and found the
vibrational entropy of disordering to be at least $0.19k_{B}$. This finding
was corroborated with subsequent incoherent neutron scattering measurements 
\cite{fultz:ni3alphon}, which bracketed its value between $0.1k_{B}$ and $%
0.3k_{B}$. These findings fueled much of the interest of the recent
theoretical literature on the Ni$_{3}$Al compound \cite
{ackland:vol,althoff:vib,althoff:vibcms,ravelo:vib,avdw:ni3al}.
Unfortunately, the agreement between experimental and theoretical
determinations is relatively poor.
Even among studies in which the magnitude of the vibrational
entropy is similar, its proposed physical origin differs
substantially: according to experiments, the vibrational entropy change occured
with essentially no change in volume, while most calculations
\cite{althoff:vib,althoff:vibcms,ravelo:vib}\ attribute the vibrational entropy 
change almost entirely to a volume change. Many of these conflicting findings were clarified by
first-principles calculations \cite{avdw:ni3al}, which predicted both
a very small volume change and a very small vibrational entropy change upon disordering.
These results indicated that
(i) the apparent large vibrational entropy change observed experimentally could very
well be entirely explained by the nanocrystalline nature of the samples and the
use of the virtual crystal approximation in interpreting the data and
(ii) the large volume expansion upon
disordering predicted by the EAM is probably an artifact of the method, as it would have otherwise
been clearly visible experimentally.
To be fair, Fultz {\it et al.} and Althoff {\it et al.} were fully aware of these potential problems;
{\it Ab initio} calculations merely made it possible to better quantify the errors introduced.
The general consensus among researchers is now that given the
numerous difficulties faced when studying Ni$_{3}$Al, unambiguous evidence
of the importance of lattice vibrations should probably be sought in other
systems.

Similar calorimetry measurements were then performed on the Fe$_{3}$Al \cite
{fultz:fe3al} and Cu$_{3}$Au \cite{fultz:cu3au} compounds, where more
conclusive results could be obtained. The vibrational entropy of disordering
of Fe$_{3}$Al was determined to be $0.10k_{B}$, a result which was later
corroborated by calculations \cite{shaojun:feal}. In the case of Cu$_{3}$Al
the experimental result, $(0.14 \pm 0.05) k_{B}$, showed very good agreement with the
earlier theoretical predictions of $0.12k_{B}$ \cite{cleri:cu3au}.
Subsequent linear-response calculations \cite{ozolins:cuau} yield values
ranging from $0.06$ to $0.08 k_{B}$ (depending on temperature), which 
also compares favorably with the experimental results.

The estimation of vibrational effects was also addressed by directly probing
the lattice dynamics through neutron scattering measurements. The
vibrational entropy of formation of a disordered alloy in the Fe-Cr system
was obtained from single crystal measurements of phonon dispersion curves 
\cite{fultz:fecr} in the virtual crystal approximation. Values ranging from
$0.14k_{B}$ to $0.21k_{B}$ were obtained, depending on concentration.

In order to determine the lattice dynamics of disordered alloys beyond the
virtual crystal approximation, the incoherent neutron scattering technique
was extensively used and refined \cite
{fultz:ni3v,fultz:co3v,fultz:co3v2,fultz:cu3au2,fultz:ni3alphon}. With this
technique, the analysis of the experimental data is considerably simplified
when the species present have comparable incoherent neutron scattering
intensities, which lead to the study of two compounds satisfying this
requirement: Ni$_{3}$V and Co$_{3}$V. Measurements on the Ni$_{3}$V compound 
\cite{fultz:ni3v} found a surprisingly small vibrational entropy change upon
disordering, $0.04k_{B}$, while the Co$_{3}$V compounds exhibited a
relatively large value $0.15k_{B}$ \cite{fultz:co3v}. A related study found
the vibrational entropy change associated with the fcc-hcp transition of the
Co$_{3}$V compound to be $0.07k_{B}$ \cite{fultz:co3v2}. It is interesting
to note that the disordering reaction exhibits a larger vibrational entropy
change than the allotropic transformation in Co$_{3}$V. Perhaps more
importantly, the investigations of the Ni$_{3}$V and Co$_{3}$V compounds
presented the first experimental evidence of important anharmonic effects.

The same technique of incoherent neutron scattering was employed to revisit
the Cu-Au system \cite{fultz:cu3au2}. The vibrational entropy of formation
of the Cu$_{3}$Au compound was found to be $0.06k_{B}$ at $300K$,
corroborating earlier estimations based on phonon dispersion curve
measurements \cite{fultz:trends}. At $800K$, the measured value of
$ (0.12 \pm 0.04) k_{B} $ is consistent with the value of $0.20k_{B}$
obtained with ab-initio calculations \cite{ozolins:cuau}.

The vibrational entropy of formation of various ordered compounds, obtained
from single crystal phonon dispersion measurements, were recently compiled 
\cite{fultz:trends} and show formation values of up to $0.5k_{B}$. However,
this compilation contains many systems where the alloy has a crystal
structure that differs from the one of the pure elements and the formation
values thus also include the vibrational entropy change associated with a
structural transition. When all these cases are excluded, the maximum
vibrational entropy change decreases to a more conservative upper bound of 
$0.20k_{B}$, which is reached in the ordered phase of Ni$_{3}$Al. 

\subsection{Conclusion}

Although early investigations of the impact of vibrational effects on phase
stability consistently found large effects, it is now becoming apparent, as
more precise theoretical and experimental techniques became available, that
vibrational effects are often, but not always, large. It is therefore important
to identify the factors which determine when they are, so that the effort
devoted to calculating them is proportional to their expected magnitude.

\section{The origin of vibrational entropy differences between phases}

\label{secmecha}We have presented the framework that allows for the
inclusion of vibrational effects in phase diagram calculations. However, the
formalism presented so far does not directly provide any intuition regarding
the origin of vibrational entropy differences. This intuition is important
to be able the predict when vibrational effects should be important and,
when they are, which approximation should be used to calculate them.

Three mechanisms have been suggested to explain the origin of vibrational
entropy differences in alloys. We will discuss them in turn.

\subsection{The ``bond proportion'' effect}

In most theoretical studies based on simple models systems \cite
{matthew:latvib,bakker:oddsvib,bakker:isingvib,tuijn:odheat,gdg:vib1}, the
effect of the state of order of an alloy on its vibrational entropy has been
attributed to the fact that bonds between different chemical species have a
different stiffness than the bonds between identical species. When the
proportion of each type of bond in the alloy changes during, for instance,
an order-disorder transition, the average stiffness of the alloy changes as
well, resulting in a change of its vibrational entropy. This so-called
``bond-proportion'' mechanism is illustrated in Fig. \ref{bondprop} in the
case of an order-disorder transition. In a system with ordering tendencies,
the bond between unlike atoms are associated with an increased stability and
are thus expected to be stiffer than bonds between alike atoms. Since
disordering reduces the number of bonds between unlike atoms in favor of
bonds between similar atoms, the disordered state is expected to be softer,
and thus, have a large vibrational entropy. Vibrations would then tend to
stabilize the disordered state relative to the ordered state, reducing the
transition temperature. A similar reasoning in the case of a phase
separating system shows that the disordered state should be softer than a
phase separated mixture, indicating that the miscibility gap should be
lowered as a result of vibrational effects.

The presence of a ``bond proportion'' effect can be readily identified from
the nature of the changes taking place in the phonon densities of states
during an order-disorder transition. In the ordered alloy, the very stiff
nearest-neighbor bonds should be associated with high frequency optical
modes peaks. As the alloy disorders, the height of these peaks should
decrease since the number of stiff bonds decreases. This characteristic
signature of the ``bond proportion'' mechanism in the phonon DOS has been
repeatedly observed in experiments \cite{fultz:ni3alphon,fultz:fe3al}
as well as in theoretical calculations \cite{althoff:vib} (although, in
the latter study the ``bond proportion'' mechanism was dominated by other
effects discussed below).

\subsection{The volume effect}

It is well known that vibrational entropy of a given compound varies with
volume: This dependence is responsible for thermal expansion. It is thus
expected that the volume change that typically occurs during solid-state
transitions should also result in a change in vibrational entropy. As the
alloy expands (or contracts), as a result of a change in its state of order,
the stiffness of all chemical bonds decreases (or increases). The resulting
change in vibrational entropy is entirely due to anharmonicity, in contrast
to the ``bond proportion'' effect. When the volume mechanism operates
alone, the phonon DOS should exhibit an overall shift when the volume
changes (See Fig. \ref{volmech}).

This shift is usually accompanied by a change in the shape of the phonon
DOS, so that visual inspection of the phonon DOS is usually not sufficient
to identify the shift due to the effect of volume. Theoretical investigations of this effect
have thus relied on a simple thought experiment consisting in separating the
vibrational entropy change upon disordering at constant volume from the
vibrational entropy change resulting solely from the volume expansion of the
disordered state. In this fashion, the importance of volume changes was
first observed in embedded atom method (EAM) calculations of disordering
reaction of the Ni$_{3}$Al and Cu$_{3}$Au compounds \cite{ackland:vol} and
later corroborated by subsequent EAM calculations on the Ni$_{3}$Al compound 
\cite{althoff:vib,ravelo:vib}. These more recent calculations also found
that the volume effect is magnified by the fact that the linear thermal
expansion coefficient of different phases can differ substantially (by about 
$5\times 10^{-6}$ $K^{-1}$ relative to an absolute value of about $15\times
10^{-6}$ $K^{-1}$). First-principles calculations on the Cu-Au system \cite
{ozolins:cuau} revealed a similar finding. In contrast, first-principles
calculations on the Ni$_{3}$Al \cite{avdw:ni3al} and Pd$_{3}$V \cite
{avdw:pd3v}\ compounds found only a small difference between the thermal
expansion coefficients of the ordered and the disordered phases (less than $%
1\times 10^{-6}$ $K^{-1}$).

Even though the temperature-dependence of vibrational entropy can be large,
it is important to keep in mind that a given vibrational entropy difference
arising from anharmonic contributions has a smaller impact on the
vibrational free energy than a harmonic contribution of the same magnitude.
The reason is that anharmonic contributions to the vibrational entropy are
always partly canceled by anharmonic contributions to the vibrational {\em %
enthalpy}. In contrast, such cancellation does not occur for harmonic
contributions.\footnote{%
A temperature-dependence of vibrational entropy necessarily introduces a
temperature-dependence of the vibrational enthalpy, as a consequence of the
following thermodynamic relation: $\frac{\partial H_{vib}}{\partial T}=T%
\frac{\partial S_{vib}}{\partial T}.$ The anharmonic vibrational free energy
is then a sum of two competing contributions: $F_{vib}=H_{vib}-TS_{vib}$. In
contrast, harmonic contributions to the vibrational enthalpy are
configuration-independent in the high-temperature limit, by the
equipartition theorem, and give no net contribution to vibrational free
energy differences. Vibrational entropy differences originating from
harmonic contributions thus enter the expression for vibrational free energy
differences directly, without any partial cancellation from the enthalpic
term.} The quasiharmonic approximation predicts that
anharmonic contributions have exactly half the effect of harmonic
contributions (see \cite{ozolins:cuau} or Appendix \ref{anhapp}) when the volume dependence
of the energy is quadratic while the volume dependence of the vibrational entropy is linear.

Experimental measurements have not identified the volume mechanism as a
major source of entropy differences since the simple thought experiment that
allows its identification in calculations cannot be performed
experimentally. The effect of thermal expansion on the phonon DOS, however,
is clearly seen experimentally \cite{fultz:ni3v,fultz:co3v}, due to the fact
that thermal expansion causes shifts in the phonon DOS that are not
accompanied by substantial changes in its shape. Measurements on Ni$_{3}$V 
\cite{fultz:ni3v}, Co$_{3}$V \cite{fultz:co3v} and Cu$_{3}$Au \cite
{fultz:cu3au2} all show that anharmonic contributions are not negligible.

\subsection{The size mismatch effect}

The third advocated source of vibrational entropy changes is the effect of
atomic size mismatch.\ When atoms of different sizes are constrained to
coexist on a lattice, the atoms can experience compressive (or tensile)
stress that results in locally stiffer (or softer) regions. When large atoms
sit on neighboring lattice sites, the amplitude of their vibrations is
reduced, {\it i.e.}, the alloy tends to be locally stiffer. Conversely, when
small atoms sit on neighboring lattice sites, the extra room available
results in a locally softer region.

The phenomenon was first noted in EAM calculations on the Cu$_{3}$Au
compound \cite{ackland:vol}, where, in the disordered state, the presence of
highly compressed pairs of Au atoms lowers the vibrational entropy of the
disordered state. A similar effect was found in first-principles
calculations on Ni$_{3}$Al \cite{avdw:ni3al}, where very compressed pairs of
Al atoms where found in the disordered state. In first-principles
calculations on Pd$_{3}$V \cite{avdw:pd3v}, an even more intriguing size-related effect was
observed: all three types of chemical bonds have incompatible equilibrium
lengths and the vibrational entropy changes can be explained solely by the
large relaxations of the atoms away from their ideal lattice sites in the
disordered state. A summary of these observations can be found in \cite{morgan:aruba}.

In the experimental literature, the size mismatch effect has been described
with the help of a stiff sphere picture first introduced in an investigation
of the Cu$_{3}$Au compound \cite{fultz:cu3au}. The fundamental intuition
behind this picture is that a chemical bond becomes stiffer when the two
bonded atoms have touching atomic ``spheres''. Further evidence for this
stiff sphere picture was provided by a systematic analysis of the
vibrational entropy of formation of various L1$_{2}$ compounds \cite
{fultz:trends}, which shows a correlation with the difference in radii
between the two alloyed species.

\section{Computational techniques}

\label{compute}To understand the nature of the difficulties encountered, it
is instructive to first consider how, in principle, the vibrational
properties of a single configuration ${\bf\sigma}$ can be calculated with
an arbitrary accuracy. The techniques presented in this section are the
tools that were used to investigate the importance of lattice vibrations
presented in Section \ref{assesscalc}.

Phase diagram calculation involves computing vibrational properties for a
set of configurations ${\bf\sigma}$. Carrying out the full phonon problem
for each configuration results in undue computational requirements.
Nevertheless the formal solutions presented here play an important role in
devising practical ways to include vibrational effects in phase diagram
calculations. This section first focuses on the treatment lattice vibration
within the harmonic approximation, before addressing the issue of
anharmonicity. Finally, important consideration concerning the energy models
used as an input for these procedures are discussed.

\subsection{Lattice vibrations in the harmonic approximation}

\label{bvk}In this section, we review the problem of determining the
constrained free energy of a system in the neighborhood of a configuration $%
{\bf\sigma}$, under the assumption that the system spends most of its time
in a region near a local energy minimum, where a harmonic approximation to
the energy surface is accurate. In this approximation, the free energy
determination reduces to the well-known phonon problem \cite{maradudin:harm}%
, \cite{ashmer:sol}.

\subsubsection{Theory}

Consider a system consisting of $N$ atoms. Let $M_{i}$ be the mass of atom $%
i $ and $u\left( i\right) $ be its displacement away from its equilibrium
positions. Time derivatives are denoted by dots while Greek letter
subscripts denote one of the cartesian components of a vector. In the
harmonic approximation, the energy of the system can be written as: 
\begin{equation}
H=\frac{1}{2}\sum_{i}M_{i}\left( \dot{u}\left( i\right) \right) ^{2}+\frac{1%
}{2}\sum_{i,j}u^{T}\left( i\right) \Phi \left( i,j\right) u\left( j\right) 
\end{equation}
where 
\begin{equation}
\Phi _{\alpha \beta }\left( i,j\right) =\left. \frac{\partial ^{2}E}{%
\partial u_{\alpha }\left( i\right) \partial u_{\beta }\left( j\right) }%
\right| _{u\left( l\right) =0\text{ }\forall l}. 
\end{equation}
The $3\times 3$ matrices $\Phi \left( i,j\right) $ are called the force
constants tensors, as they relate the displacement of atom $j$ to the force $%
f$ exerted on atom $i$: 
\begin{equation}
f\left( i\right) =\Phi \left( i,j\right) u\left( j\right) . 
\end{equation}
Such a harmonic approximation of a solid is often referred to as a Born-von
K\'{a}rm\'{a}n model.

Note that contrary to usual treatment, we do not immediately impose
translational symmetry, in order to derive a few general results that also
apply to systems such as disordered alloys.

The substitution $e\left( i\right) =\sqrt{M_{i}}u\left( i\right) $ yields: 
\begin{equation}
H=\frac{1}{2}\left( \sum_{i}\dot{e}^{2}\left( i\right)
+\sum_{i,j}e^{T}\left( i\right) \frac{\Phi \left( i,j\right) }{\sqrt{%
M_{i}M_{j}}}e\left( j\right) \right) .  \label{Heharm}
\end{equation}
The $3N$ eigenvalues $\lambda _{m}$ of the matrix\footnote{%
Among the $3N$ eigenvalues, the $6$ eigenvalues associated with rigid body
translations and rotations are zero. In the thermodynamic limit, these few
degrees of freedom are inconsequential. To avoid notational complications,
we simply assume that the solid is fixed to a reference frame by springs so
that the resulting dynamical matrix has $3N$ nonzero eigenvalues.} 
\begin{equation}
D=\left( 
\begin{array}{ccc}
\frac{\Phi \left( 1,1\right) }{\sqrt{M_{1}M_{1}}} & \cdots & \frac{\Phi
\left( 1,N\right) }{\sqrt{M_{1}M_{N}}} \\ 
\vdots & \ddots & \vdots \\ 
\frac{\Phi \left( N,1\right) }{\sqrt{M_{N}M_{1}}} & \cdots & \frac{\Phi
\left( N,N\right) }{\sqrt{M_{N}M_{N}}}
\end{array}
\right) .  \label{bigdynmat}
\end{equation}
then give the frequencies $\nu _{m}=\frac{1}{2\pi }\sqrt{\lambda _{m}}$ of
the normal modes of oscillation. In the harmonic approximation, the
knowledge of these frequencies is sufficient to determine the thermodynamic
quantities we are interested in. This information is conveniently summarized
by the phonon density of states (DOS), which gives the number of modes of
oscillation having a frequency lying in the interval $\left[ \nu ,\nu +d\nu 
\right] $: 
\begin{equation}
g\left( \nu \right) =\frac{1}{N}\sum_{m=1}^{3N}\delta \left( \nu -\nu
_{m}\right) . 
\end{equation}
It can be shown that the free energy of the system (restricted to remain
close to a given configuration ${\bf\sigma}$) is given by \cite
{maradudin:harm}: 
\begin{eqnarray*}
\frac{F}{N} &=&\frac{E^{\ast }}{N}+k_{B}T\int_{0}^{\infty }\ln \left( 2\sinh
\left( \frac{h\nu }{2k_{B}T}\right) \right) g\left( \nu \right) d\nu \\
&=&\frac{E^{\ast }}{N}+\frac{k_{B}T}{N}\sum_{m}\ln \left( 2\sinh \left( 
\frac{h\nu _{m}}{2k_{B}T}\right) \right)
\end{eqnarray*}
where $E^{\ast }$ is the potential energy of the system at its equilibrium
position and $h$ is Planck's constant. Phase transitions in alloys typically
occur at a temperature where the high temperature limit of this expression
is an accurate approximation:
\begin{eqnarray*}
\frac{F}{N} &=&\frac{E^{\ast }}{N}+k_{B}T\int_{0}^{\infty }\ln \left( \frac{%
h\nu }{k_{B}T}\right) g\left( \nu \right) d\nu \\
&=&\frac{E^{\ast }}{N}+\frac{k_{B}T}{N}\sum_{m}\ln \left( \frac{h\nu _{m}}{%
k_{B}T}\right)
\end{eqnarray*}
The usual criterion used to determine the temperature range where high
temperature limit is reached is the Debye temperature. Note that the factor $%
\frac{h}{k_{B}T}$ is often omitted because it cancels out when calculating
vibrational free energy differences. In the high temperature limit, another
important form of cancellation occurs: The atomic masses have no effect on
the free energies of formation \cite{grimvall:nomass,gdg:vib1}. This important result, shown in Appendix \ref
{masscancel}, rules out that masses play any significant role in determining
phase stability at high temperatures.

As mentioned before, a convenient measure of the magnitude of the effect of
lattice vibrations on phase stability is the vibrational entropy, which can
be obtained from the vibrational free energy by the well known
thermodynamical relationship $S_{\text{vib}}=-\frac{\partial F_{\text{vib}}}{%
\partial T}$. Contrary to the vibrational free energy of formation, the
vibrational entropy of formation\footnote{%
The absolute value of the vibrational entropy is not constant at high
temperature, but its temperature-dependence does not vary across distinct
phases and thus formation values are temperature-independent.} is
temperature-independent in the high-temperature limit of the harmonic
approximation, allowing a unique number to be reported as a measure of the
importance of vibrational effects.

In a crystal, the determination of the normal modes is somewhat simplified
by the translational symmetry of the system. Let $n$ denote the number of
atoms per unit cell. Let $u\binom{l}{i}$ denote the displacements away from
its equilibrium position of atom $i$ in cell $l$. Let $\Phi \binom{l\text{ }%
l^{\prime }}{i\text{ }j}$ be the force constant relative to atom $i$ in cell 
$l$ and atom $j$ in cell $l^{\prime }$ and let $e\binom{l}{i}=\sqrt{M_{i}}u%
\binom{l}{i}$. Bloch's theorem indicates that the eigenvectors of the
dynamical matrix are of the form 
\begin{equation}
e\binom{l}{i}=e^{\iota 2\pi \left( k\cdot l\right) }e\binom{0}{i}
\label{eperio}
\end{equation}
where $l$ denotes the cartesian coordinates of one corner of cell $l$ and $k$
is a point in the first Brillouin zone. This fact reduces the problem of
diagonalizing the $3N\times 3N$ matrix $D$ to the problem of diagonalizing a 
$3n\times 3n$ matrix $D\left( k\right) $ for various values of $k$. This can
be shown by a simple substitution of Equation (\ref{eperio}) into Equation (%
\ref{Heharm}).\footnote{%
And changing the summation over atoms by summations over atoms and cells.}\
The dynamical matrix $D\left( k\right) $ to be diagonalized is given by%
\footnote{%
The reader should be aware that there are many possible conventions regarding
the phase factor: for instance, $e^{i2\pi \left( k\cdot l\right) }$, $%
e^{-i2\pi \left( k\cdot l\right) }$, $e^{i\left( k\cdot l\right) }$, $%
e^{i2\pi \left( k\cdot \left( l+x\left( j\right) \right) \right) }$ where $%
x\left( j\right) $ is the coordinate of atom $j$ within the cell. While all
convention yield different dynamical matrices, they all have the same
eigenvalues.} 
\begin{equation}
D\left( k\right) =\sum_{l}e^{i2\pi \left( k\cdot l\right) }\left( 
\begin{array}{ccc}
\frac{\Phi \binom{0\,l}{1\,1}}{\sqrt{M_{1}M_{1}}} & \cdots & \frac{\Phi 
\binom{0\,l}{1\,n}}{\sqrt{M_{1}M_{n}}} \\ 
\vdots & \ddots & \vdots \\ 
\frac{\Phi \binom{0\,l}{n\,1}}{\sqrt{M_{n}M_{1}}} & \cdots & \frac{\Phi 
\binom{0\,l}{n\,n}}{\sqrt{M_{n}M_{n}}}
\end{array}
\right) . 
\end{equation}
As before, the resulting eigenvalues $\lambda _{i}\left( k\right) $ for $%
i=1\ldots n$, give the frequencies of the normal modes ($\nu _{i}\left(
k\right) $ =$\frac{1}{2\pi }\sqrt{\lambda _{i}\left( k\right) }$). The
function $\nu _{i}\left( k\right) $ for a given $i$ is called a phonon
branch, while the plot of the $k$-dependence of all branches along a given
direction in $k$ space is called the phonon dispersion curve. In periodic
systems, the phonon DOS is defined as 
\begin{equation}
g\left( \nu \right) =\sum_{i=1}^{3n}\int_{BZ}\delta \left( \nu -\nu
_{i}\left( k\right) \right) dk 
\end{equation}
where the integral is taken over the first Brillouin zone.

\subsubsection{Force Constant Determination}

\label{findfk}

The above theory relies, of course, on the availability of the force
constant tensors. The determination of these force constant tensors is the
focus of this section. Before describing the methods used for their
determination, we will first review important properties of the force
constant tensors.

While the number of unknown force constants to be determined is in principle
infinite, it can, in practice, be reduced to a manageable finite number with
the help of the following two observations. First, the force constant $\Phi
(i,j)$ between two atoms $i$ and $j$ beyond a given distance can be
neglected. Second, the symmetry of the crystal imposes linear constraints
between the elements of the force constant tensors.

The accuracy of the approximation made by truncating the range of force
constant can be tested by gradually increasing the range of interactions,
until the quantities to be determined no longer vary substantially. It is
important to note that most thermodynamic quantities can be written as a
weighted integral of the phonon DOS and their convergence rates are thus
much faster than the pointwise convergence rate of the phonon DOS itself 
\cite{gdg:phd,avdw:ni3al}. That is, the errors on the DOS at each frequency
tend to be quickly averaged out when the contributions of each frequency are
added.

The restrictions on the force constants imposed by the symmetry of the
lattice can be expressed as follows. Consider the force constant $\Phi (i,j)$
of atoms $i$ and $j$ located at $x(i)$ and $x(j)$ and consider a symmetry
transformation that maps a point of coordinate $x$ to $Sx+t$, where $S$ is a 
$3\times 3$ matrix and $t$ and $3\times 1$ translation vector. In general,
if the crystal is left unchanged by such a symmetry operation, the force
constant tensors should be left unchanged as well. This fact imposes the
following constraints on the spring tensors: 
\begin{equation}
\left. 
\begin{array}{c}
Sx(i)+t=x(i^{\prime }) \\ 
Sx(j)+t=x(j^{\prime })
\end{array}
\right\} \Rightarrow \Phi (i,j)=S^{T}\Phi (i^{\prime },j^{\prime })S\text{.} 
\end{equation}
Additional constraints on the force constants can be derived from simple
invariance arguments. The most important constraints, obtained by noting
that rigid translations and rotations must leave the forces exerted on the
atoms unchanged, are 
\begin{eqnarray*}
\Phi \left( i,i\right) &=&-\sum_{j\not=i}\Phi \left( i,j\right) \\
\Phi \left( i,j\right) &=&\Phi ^{T}\left( j,i\right)
\end{eqnarray*}
Additional constrains can be found in \cite{maradudin:harm,born:phonon}.

There are essentially three approaches to determining the force constants:
analytic calculations, supercell calculations and linear response
calculations. Analytic calculations are only possible when the energy model
is sufficiently simple to allow a direct calculation of the second
derivatives of the energy with respect to atomic displacements, as in the
case of empirical pair potential models. For first-principles calculations,
either one of the two following methods have to be used.

\paragraph{The supercell method}

The supercell method \cite{wei:vib},\cite{gdg:phd}\ consists of slightly
perturbing the positions of the atoms away from their equilibrium position
and calculating the reaction forces. Equating the calculated forces to the
forces predicted from the harmonic model yields a set of linear constraints
that allows the unknown force constants to be determined.\footnote{%
Equalities between calculated and predicted energies can be used as well.
Using energies alone to determine the force constants would be a rather
inefficient use of the information provided by ab-initio calculations. Once
a first-principles calculations of the energy of a distorted structure has
been completed, the calculation of the forces acting on the atoms is
computationally inexpensive. The knowledge of the energy provides a single
equation while the knowledge of the forces provide up to $3$ equations per
atom.} 
\begin{equation}
F(i)=\Phi (i,j)u(j) 
\end{equation}

When the force constants considered have a range that exceeds the extent of
the primitive cell, a supercell of the primitive cell has to be used. (The
simultaneous movement of the image atoms introduces linear constrains among
the forces that prevent the determination of some of the force constants.)

While any choice of the perturbations that allows the force constants to be
determined is in principle equally valid, a few simple principles
drastically narrow down the number of perturbations that need to be
considered. For a given supercell, there is a only of finite number of
non-redundant perturbations to consider.

A minimal set of non-redundant perturbations can be obtained as follows.

\begin{itemize}
\item  Consider in turn each atom in the asymmetric unit of the primitive
cell.

\item  Mark the chosen atom (and its periodic images in the other
supercells) and consider it as distinct from other atoms of the same type.
(This operation effectively removes some of the symmetry operation of the
space group of the crystal.)

\item  Construct the point group $\left\{ S_{i}\right\} $ of the site where
this atom is located. ($S_{i}$ is a $3\times 3$ matrix.)

\item  Move the chosen atom along a direction $u_{1}$ such that the space
spanned by the vectors $S_{i}u_{1}$ (for all $i$) has the highest
dimensionality possible.

\item  If the resulting dimensionality is less than three, consider an
additional direction $u_{2}$ such that the space spanned by the vectors $%
S_{i}u_{j}$ for $j=1,2$ has the highest dimensionality possible.

\item  If the resulting dimensionality is less than three, consider a
direction $u_{3}$ orthogonal to $u_{1}$ and $u_{2}$.
\end{itemize}

The resulting displacements $u_{j}$ for all atoms in the asymmetric unit
gives a minimal list of perturbations that is sufficient to find all the
force constants that can possibly be determined with the given supercell.
This result follows from the observation that any other possible
displacement can be written as a linear combination of the displacements
considered above (or displacements that are symmetrically equivalent to
them).

When determining force constants with the supercell method, it is important
to verify that the presence of small numerical noise in the calculated
forces does not result in too much error in the fitted force constants. To
minimize noise in the fitted force constants, it may be necessary to use
more than the minimum possible number of perturbations. The additional
perturbations should ideally be based on different supercells, to minimize
the systematic errors introduced by the movement of the image atoms.

When ab-initio calculations are used to calculate the forces, it is
especially important to iterate the electronic self-consistency steps to
convergence. Even though the energy may appear to be well converged, the
forces may not yet be. Energy is the solution to a minimization procedure,
while forces are not. As a result, errors on the energy are of a second
order in the minimization parameters, while the errors on the forces are of
the first order in the minimization parameters. For the same reason, special
attention should be given to the structural relaxations.

The true system is not exactly harmonic and the calculated forces may
exhibit anharmonic components that introduce noise into the fitted force
constants. This problem can be alleviated by considering an additional set
of perturbations, where the displacements have the opposite sign.
Subtracting the calculated forces obtained for this new set of displacements
from the corresponding displacements of the opposite sign exactly cancels
out all the odd-order anharmonic terms. Of course, for perturbations such
that the negative displacement is equivalent by symmetry to the
corresponding positive displacement, this duplication is unnecessary,
because the terms of odd order are already zero by symmetry.

Additional guidelines for fitting force constants can be found in \cite
{ackland:prac,wei:vib,gdg:phd}.

\paragraph{Linear response}

Linear response calculations seek to directly evaluate the dynamical matrix
for a set of $k$ points. The starting point of the linear response approach
is evaluation of the second-order change in the electronic energy induced by
a atomic displacements from perturbation theory. Within this framework,
practical schemes to compute vibrational properties in semiconductors \cite
{baroni:lr,giannozzi:lr,gonze:lr,waghmare:phd} and metallic systems \cite
{gironcoli:lr,quong:al,ozolins:phd} have been devised. In this section we
will not discuss the theory behind linear response calculations
which can be found in a recent review \cite{gonze:lr_rev1,gonze:lr_rev2},
but rather
focus on how the results of linear response calculations can be used in the
context of alloy phase diagram calculations.

The dynamical matrices calculated from linear response theory are exact in
the sense that they account for arbitrarily long-range force constants.
While in the supercell method inaccuracies arise from the truncation of the
force constants, the limit in precision for linear response calculations
arises from the use of a small number of $k$ points to sample the Brillouin
zone. To address this issue, two methods can be used.

A set of special $k$ points can be selected through the Chadi-Cohen \cite
{chadi:specialk} or Mon{\-}khorst-Pack \cite{monkhorst:specialk} schemes.
Special $k$ points are selected so that the integral over the Brillouin zone
of a function $h\left( k\right) $ that contains no Fourier components above
a given frequency can be exactly evaluated by a weighted average of the
function at each special point. Since thermodynamic quantities can be
written as integrals of functions of the dynamical matrix $f\left( D\left(
k\right) \right) $ over the Brillouin zone, the procedure is straightforward
to apply in this context.

The other approach is the so-called Fourier inversion method (see, for
instance, \cite{giannozzi:lr,quong:al}). The calculated dynamical matrices
from a set of $k$ points are used to determine the value of the force
constants up to a certain interaction range. The resulting harmonic model
can then be used to calculate the dynamical matrix at any point in the
Brillouin zone, allowing a much finer sampling of the Brillouin zone for the
purpose of performing the numerical integration required to determine any
thermodynamic quantity.

The Fourier inversion method is preferable when the function $f\left(
D\left( k\right) \right) $ to be integrated exhibits high-frequency
components, while the dynamical matrix itself, $D\left( k\right) $, does
not. Such a situation would arise when the function $f$ is highly
nonlinear. The smoothness of $D\left( k\right) $ then ensures that it can be
represented with a small number of Fourier components. The less well-behaved
function $f\left( D\left( k\right) \right) $ can then be accurately
integrated with as many $k$ points as needed, using the dynamical matrix $%
D\left( k\right) $ calculated from the spring model.

In the case of vibrational free energy calculations, the special $k$ points
method has been observed to converge rapidly with respect to the number of $%
k $ points \cite{gdg:vib2}\footnote{%
Calculations of the authors, based on data from \cite{ozolins:cuau} also
support this finding.}, so that the Fourier inversion method is probably
unnecessary.\footnote{%
Note that the function whose integral gives the free energy exhibits a
logarithmic singularity at the $\Gamma $ point, which could lead to high
frequency components that are difficult to integrate accurately. However, in
three-dimensional systems, this logarithmic singularity contributes very
little to the value of the free energy, so that the rate of convergence of
the integral as a function of the number of $k$ points is not dramatically
slowed down by the presence of the singularity. As a result, the special $k$
point method can safely be used in practical calculations of the vibrational
free energy.}

For a given set of special $k$ points, there is an approximate
correspondence between the number of Fourier components that can be
integrated exactly and the range of force constants that can be determined.
The correspondence is exact only when the lattice has one atom per cell and
when the function $f$ is linear.\footnote{%
As can be shown by a simple Fourier transform.}

While supercell and linear response calculations are in principle equivalent
in terms of the information they provide, they have complementary advantages
in terms of computational efficiency. The linear response method is the most
efficient way to perform high-accuracy calculations that would otherwise be
tedious and computer intensive with the supercell method. However, when a
high accuracy is not needed, the supercell method has the advantage that
various simplifying assumptions regarding the structure of the force
constant tensors can transparently be used to drastically reduce
computational requirements. It is not clear at which level of accuracy the
cross-over between the efficiency of each approach occurs, but it is
important to keep both approaches in mind. Another consideration is that in
the continuously evolving field of computational solid state physics, new
first-principles energy methods are continually developed, and the
derivation of the appropriate linear response theory always follows the
derivation of simple force calculations. Hence, despite the elegance of
linear response theory, it is to be expected that the supercell method will
always remain of interest.

\subsection{Anharmonicity}

While the harmonic approximation is remarkably accurate given its
simplicity, it has one important limitation: It is unable to model thermal
expansion and its impact on vibrational properties. Both the free energy $F$
and the entropy $S$ can be obtained from the the heat capacity $C_{p}$: 
\begin{equation}
S=\int_{0}^{T}\frac{C_{p}}{T}dT\text{ and }F=\left. E\right|
_{T=0}-\int_{0}^{T}S\,dT, 
\end{equation}
Hence, a simple way to account for thermal expansion is to use the following
well known thermodynamic relationship between the heat capacity at constant
pressure $C_{p}$ and at constant volume $C_{v}$: 
\begin{equation}
C_{p}=C_{v}+BVT\alpha ^{2}  \label{cpcv}
\end{equation}
where $\alpha $ is the coefficient of volumetric thermal expansion while $B$
is the bulk modulus. In a purely harmonic model, there is no thermal
expansion and $C_{v}$ is equal to $C_{p}$. 
The term $BVT\alpha ^{2}$ can thus be viewed as correction arising from anharmonic
effects.

Equation (\ref{cpcv}) is directly useful in the context of experimental
measurements where $C_p$, $V$ and $\alpha$ can be directly measured \cite
{fultz:ni3v,fultz:co3v2}. In the following section, we describe the
computational techniques used to handle anharmonicity.

\subsubsection{The quasiharmonic model}

A simple modification to the harmonic approximation, called the
quasiharmonic approximation, allows the calculation of thermal expansion at
the expense of a moderate increase in computational cost. In the
quasi-harmonic approximation, the phonon frequencies are allowed to be
volume-dependent, which amounts to assuming that the force constant tensors
are volume-dependent (see, for instance, \cite{grimvall:thermophys}).
This approximation has recently been shown
to be extremely reliable, enabling accurate first-principles calculations
of the thermal expansion coefficients of many elements up to their melting points
\cite{quong:thexp}.

The best way to understand this approximation is to
study a simple model system where it is essentially exact. Consider a linear
chain (with periodic boundary conditions) of identical atoms interacting
solely with their nearest neighbors through a pair potential of the form: 
\begin{equation}
U\left( r\right) =a_{1}r+a_{2}r^{2}+a_{3}r^{3}. 
\end{equation}
Let $L$ be the average distance between two nearest neighbors and let $%
u\left( i\right) $ denote the displacement of atom $i$ away from its
equilibrium position. The total potential energy (per atom) of this system
is then given by 
\begin{eqnarray*}
\frac{U}{N}&=&\frac{1}{N}\sum_{i}a_{1}\left( L+u\left( i\right) -u\left(
i+1\right) \right) +a_{2}\left( L+u\left( i\right) -u\left( i+1\right)
\right) ^{2} \\
& &+a_{3}\left( L+u\left( i\right) -u\left( i+1\right) \right) ^{3}
\end{eqnarray*}
This expression can be simplified by noting that all the terms that are
linear in $\left( u\left( i\right) -u\left( i+1\right) \right) $ cancel out
when summed over $i$. 
\begin{eqnarray*}
\frac{U}{N}&=&a_{1}L+a_{2}L^{2}+a_{3}L^{3}+\left( a_{2}+3a_{3}L\right) \frac{%
1}{N}\sum_{i}\left( u\left( i\right) -u\left( i+1\right) \right) ^{2} \\
& &+O\left(\left( u\left( i\right) -u\left( i+1\right) \right) ^{3}\right)
\end{eqnarray*}
The first three terms, $a_{1}L+a_{2}L^{2}+a_{3}L^{3}$, give the elastic
energy of a motionless lattice while the remaining terms account for lattice
vibrations. The important feature of this equation is that, even within the
harmonic approximation, the prefactor of the harmonic term, $\left(
a_{2}+3a_{3}L\right) $, depends on the anharmonicity of the potential
(through $a_{3}L$). In the more realistic case of three-dimensional systems,
this length-dependence translates into a volume-dependence\footnote{%
Of course, in general, it could be a general strain dependence, if the
symmetry of the crystal is sufficiently low.} of the harmonic force
constants $\Phi \binom{l\,l^{\prime }}{i\,j}$.

The volume dependence of the phonon frequencies induced by the volume-depen{%
\-}den{\-}ce of the force constants is traditionally modeled by the
Gr\"{u}neisen parameter 
\begin{equation}
\gamma _{kj}=-\frac{\partial \ln \nu _{j}\left( k\right) }{\partial \ln V} 
\end{equation}
which is defined for each branch $j$ and each point $k$ in the first
Brillouin zone. But since we are interested in determining the free energy
of a system, it is convenient to directly parametrize the volume dependence
of the free energy itself. This dependence has two sources: the change in
entropy due to the change in the phonon frequencies and the elastic energy
change due to the expansion of the lattice: 
\begin{equation}
F\left( T,V\right) =E^{\ast }\left( V\right) +F_{H}\left( T,V\right) 
\end{equation}
where $E^{\ast }\left( V\right) $ is the energy of a motionless lattice
constrained to remain at volume $V$, while $F_{H}\left( V\right) $ is the
vibrational free energy of a harmonic system constrained to remain at volume $V$. The
equilibrium volume $V^{\ast }\left( T\right) $ at temperature $T$ is
obtained by minimizing this quantity with respect to $V$.\ The resulting
free energy at temperature $T$ is then given by $F\left( T,V^{\ast }\left(
T\right) \right) $.\footnote{%
Formally, the free energy should be determined by a sum over every possible
volume: $-k_{B}T\ln \left( \sum_{V}\exp \left( -\beta F\left( V\right)
\right) \right) $. However, since the volume is a macroscopic quantity, its
distribution can be considered a delta function and the sum reduces to a
single term: the free energy at the volume that minimizes the free energy.}

Let us consider a particular case that illustrates the effect of temperature
on the free energy, at the cost of a few reasonable assumptions. We assume
that

\begin{itemize}
\item  the elastic energy of the motionless lattice is quadratic in volume;

\item  the high temperature limit of the free energy can be used.
\end{itemize}

As shown in Appendix \ref{anhapp}, in this approximation, the volume
expansion $\Delta V$ as a function of temperature takes on a particularly
simple form: 
\begin{equation}
\frac{\Delta V}{N}=\frac{3k_{B}T\overline{\gamma }}{B} 
\end{equation}
where $\overline{\gamma }$ is an average Gr\"{u}neisen parameter: 
\begin{equation}
\overline{\gamma }=\frac{1}{3N}\sum_{m=1}^{3N}\frac{V}{\nu _{m}}\frac{%
\partial \nu _{m}}{\partial V}. 
\end{equation}
The resulting temperature dependence of the free energy is given by

\begin{equation}
\frac{F\left( T\right) }{N}=\frac{F\left( T,V_{0}\right) }{N}-\frac{\left(
3k_{B}T\overline{\gamma }\right) ^{2}}{2B\left( V_{0}/N\right) }. 
\end{equation}
These expressions provide a simple way to account for thermal expansion.
They also allow us to estimate the changes in
vibrational entropy as a function of temperature that is due to thermal expansion:
\begin{equation}
\frac{dS_{\text{vib}}}{dT}=\frac{ 9 k_B \overline{\gamma} ^2}{B(V_0/N)} k_B .
\end{equation}
In metallic alloys, this quantity is typically of the order of
$10^{-4} k_B /\text{K }$.

\subsubsection{Simulation}

There are two main simulation-based approaches to handling anharmonicity:
Monte Carlo (MC) \cite{binder:mc} and Molecular Dynamics (MD) \cite{allen:md}.
While both approaches are able to model anharmonicity at any level of
accuracy, they suffer from two limitations. First, they are computationally
demanding and therefore have, to date, been limited to simple energy models.
Second, they are unable to model quantum mechanical aspects of vibrations
and are therefore limited to the high temperature limit.\footnote{%
Monte Carlo simulations that include quantum effects are possible for
systems containing a small number of particles.} 
There is an interesting and useful complementarity between the
quasi-harmonic model and simulation techniques \cite
{althoff:quasi,morgan:phd}. Quantum effects typically become negligible in
the temperature range where strong anharmonic effects, which cannot be
modeled accurately within the quasiharmonic framework, become important.

The use of simulation techniques to determine vibrational properties
bypasses the coarse graining framework presented in section \ref{coarse}:
Both configurational and vibrational excitations are treated on the same
level. When a simple energy model provides a sufficient accuracy,
one can calculate thermodynamic properties directly from MC simulations
where both atomic displacements and
changes in chemical species are allowed during the simulation \cite{foiles:surface,Zunger:GaInP-MC}.

While a full determination of a phase diagram from simulations has, to our knowledge,
not been attempted, both MD (\cite{ravelo:vib}) and MC (\cite{althoff:quasi,morgan:phd}) have been used to
determine differences in vibrational free energy between two phases. Because
neither MD nor MC are able to provide free energies directly, a special
integration technique needs to be used. The idea is to express a
thermodynamic quantity inaccessible to MC as an integral of a quantity that 
{\em can} be obtained through MC. A simple example is the change free
energy $F$ as a function of temperature at constant pressure, which can be
derived from the Gibbs-Helmholtz relation 
\begin{equation}
F\left( T_{2}\right) =F\left( T_{1}\right) -\int_{1/T_{1}}^{1/T_{2}}E d(1/T) 
\end{equation}
where $E$ is the internal energy. Another example is the change in
entropy as a function of temperature (at constant pressure) which can be
expressed in terms of the heat capacity:
\begin{equation}
S(T_2)=S(T_1)+\int_{T_1}^{T_2} \frac{C_p}{T}dT.
\end{equation}

Often, the most computationally efficient path of integration between two states is not physically meaningful.
For instance, one can gradually change the interatomic
potentials during the course of the simulation, in order to model a change
in the configuration of the alloy, without requiring atoms to jump between
lattice sites. This task is achieved by defining an effective Hamiltonian 
\begin{equation}
H_{\lambda }=\left( 1-\lambda \right) H^{\alpha }+\lambda H^{\beta } 
\end{equation}
that gradually switches from the Hamiltonian $H_{\alpha }$ associated with
phase $\alpha $ to the Hamiltonian $H_{\beta }$ associated with phase $\beta 
$ as the switching parameter $\lambda $ goes from 0 to 1. This convenient
path of integration permits the calculation of free energy differences
between phases at a reasonable computational cost, with the help of the
following thermodynamic relation: 
\begin{equation}
F^{\beta }=F^{\alpha }+\int_{0}^{1}\left( \left\langle H_{\beta
}\right\rangle _{\lambda }-\left\langle H_{\alpha }\right\rangle _{\lambda
}\right) d\lambda 
\end{equation}
where $\left\langle H_{\alpha }\right\rangle _{\lambda }$ is the 
average of the energy calculated using Hamiltonian $H_{\alpha }$ (and
similarly for $\left\langle H_{\beta }\right\rangle _{\lambda }$.

\subsection{Energy models}

Force constants and anharmonic contributions are ultimately always derived
from an energy model. In this section, we discuss various energy models,
from empirical potential models to first-principles techniques, and the
error or bias they may introduce in the vibrational properties.

Simple pairwise potentials of functionals (such as the Embedded Atom Method)
are computationally efficient so that all vibrational properties can often
be determined without any approximations beyond the ones associated with the
specific energy model. For this reason, the use of simple energy models has
proven to be an invaluable tool to understand trends in vibrational
entropies and to test a number of approximations \cite
{gdg:vib2,althoff:quasi,morgan:local,gdg:phd,morgan:phd}.

Several potential sources of error can arise when using pair potentials or
pair functionals. The first one is that vibrational entropy is extremely
sensitive to the precise nature of the relaxations that take place in an
alloy and a simple energy model may not be able to accurately predict these
relaxations. This problem is particularly apparent when considering the wide
range of values found in the different calculations of the vibrational
entropy change upon disordering of the Ni$_{3}$Al compound \cite
{ackland:vol,althoff:vib,ravelo:vib,avdw:ni3al}. But, as shown in Table \ref
{ni3alSV}, most of the discrepancies can be explained from differences in
the predicted volume change upon disordering.

This is often aggravated by the fact that simple energy models are often not
fitted to phonon properties. The problem was noted in \cite{althoff:vib}
where the embedded atom potentials used were fitted to various structural
energies and elastic constants \cite{voter:eam}. The acoustic modes were
accurately extrapolated from the fit to the elastic constants, but the
phonon frequencies associated with the optical modes were overestimated by
about $10$\%.\footnote{%
Most of the bias in the vibrational entropy introduced by this problem
should however cancel out when taking the difference in vibrational entropy
between two phases where the same problem is present.} The question of the
accuracy of simple energy models clearly merits further attention. In this
respect, the fit of simple energy models to the results of {\it ab-initio}
calculations \cite{Zunger:GaInP-MC,shaojun:feal} offers a promising way to
include vibrational effects. 

In oxides, electronic polarization has to be included in order to correctly
model both the low frequency acoustic modes and the high frequency optical
modes. Electronic polarization in oxides can be approximated with the so-called
core and shell model \cite{dick:coreshell}.

While quantum mechanical methods are computationally more intensive, they
generally provide more accurate force constants. The most obvious error
introduced by the common Local Density Approximation (LDA) is its systematic
underprediction of lattice constants which leads to an overestimation of
elastic constants and phonon frequencies. This systematic error makes it
difficult to compare the absolute values of calculated vibrational
properties with experimental measurements. However, for the purpose of
calculating phase diagrams, this bias may be less of a concern, because
phase stability is determined by differences in free energies, and one would
expect a large part of this systematic error to cancel out.

A practical way to alleviate the LDA bias is to perform calculations at a
negative pressure such that the calculated equilibrium volume agrees with
the experimentally observed volume. As shown in \cite{avdw:lda-p}, a very
good estimate of the required negative pressure can be obtained by a
concentration-weighted average of the pressure associated with the elemental
solids. For the purpose of calculating elastic properties, this approach
appears to outperform the most popular alternative to LDA, the Generalized
Gradient Approximation (GGA).\footnote{%
Part of the success of the ``negative pressure'' LDA is due to the fact that
it uses information regarding the true experimental volume instead of being
fully ab-initio. But the knowledge of one pressure per element is a
relatively small amount of information.}

\subsection{Convergence issues}

\label{controlapp}The basic formalisms presented in Section \ref{alloyth}
and \ref{bvk} provide two natural ways to control the trade-off between
accuracy and computational requirements. In the context of alloy theory
(Section \ref{alloyth}), the range of the effective clusters interactions
included in the cluster expansion controls how accurately the
configurational dependence of vibrational properties is modeled. In the
context of the harmonic (or quasiharmonic) treatment of lattice vibrations
(Section \ref{bvk}), the range of the force constants included in the
Born-von K\'{a}rm\'{a}n model controls the accuracy of the calculated
vibrational properties for a given configuration. In principle, any desired
accuracy can be reached, given sufficient computing power, by increasing the
range of the interactions in both the cluster expansion and the Born-von
K\'{a}rm\'{a}n models. This section seeks to answer the important question
of how far these two ranges of interactions need to be pushed in order to
reach the accuracy required in a typical phase diagram calculation.

\subsubsection{Short-range force constant}

The evidence that spring models including only short-range force constants
are able to correctly model vibrational quantities comes from various
sources.

First and second nearest neighbor spring model are routinely used to fit
data obtained from neutron scattering measurement of phonon dispersion
curves \cite{fultz:ni3alphon,fultz:cu3au}. In the theoretical literature,
there have been direct studies of the convergence as a function of the range
of interaction considered. All {\it ab-initio} studies find that short-range
force constants (first or second nearest neighbor) permit an accurate
determination of thermodynamical quantities in metals \cite
{avdw:ni3al,avdw:pd3v} and group IV semiconductors \cite{gdg:phd}. It is
important to note that this rapid convergence of most thermodynamic
quantities occurs even when the pointwise convergence rate of the phonon DOS
is slow. As noted before, this property arises from the fact that
thermodynamic quantities are averages taken over all phonon modes and errors
tend to average out.

In ionic systems, the presence of long-range electrostatic interactions may
require long-range force constants. However, this electrostatic effect can
easily be modeled using pair interactions at a moderate computational cost.
Once the forces predicted from a simple electrostatic model have been
subtracted, the residual forces should be parameterizable with a short-range
spring model.

Some of the ab-initio studies of convergence have suggested additional
simplifications to force constant tensors \cite{gdg:phd,avdw:pd3v}: instead
of attempting to compute all force constants in each tensor, is it possible
to obtain reliable results by keeping only the largest terms. We now present
a hierarchy of approximations that is a formalization of these findings.

To obtain a more intuitive representation of a given force constant tensor $%
\Phi _{\alpha \beta }\left( i,j\right) $, we express it in a basis such that
the first cartesian axis is aligned along the line joining atom $i$ and $j$.
The second axis is then taken along the highest symmetry direction
orthogonal to the first axis while the third axis is chosen so obtain a
right handed orthogonal coordinate system.

In the absence of symmetry, the most general force constant tensor has 9
independent elements. The first simplification, is to neglect the three body
terms in the harmonic model of the energy ({\it e.g. }$\left( x_{\alpha
}\left( i\right) -x_{\alpha }\left( j\right) \right) \left( x_{\beta }\left(
i\right) -x_{\beta }\left( k\right) \right) $ with $\alpha \not=\beta $).
Physically, such terms arise from the deformation of the electronic cloud
surrounding atom $i$ that is caused by moving atom $j$ and that affect the
force acting on atom $k$. Clearly, for any force constant other than the
nearest neighbor, this effect is negligibly small. Even for nearest neighbor
tensors, it is the most natural contribution to neglect first. In can be
readily shown that a solid consisting only of pairwise harmonic interaction,
the tensor associated with a pair of atoms is symmetric: 
\begin{equation}
\Phi _{\alpha \beta }\left( i,j\right) =\Phi _{\beta \alpha }\left(
i,j\right) . 
\end{equation}
(This constraint is distinct from the conventional constraint: $\Phi
_{\alpha \beta }\left( i,j\right) =\Phi _{\beta \alpha }\left( j,i\right) .$)

The elements of the force constant tensor can be ranked in decreasing order
of expected magnitude based on three simple assumptions:

\begin{enumerate}
\item  Force constants associated with stretching a bond are larger than the
ones associated with bending it.

\item  Terms relating orthogonal forces and displacements are smaller than
those relating parallel forces and displacements.

\item  In the plane perpendicular to the bond, the anisotropy in the force
constants is smaller than the magnitude of the force constants themselves.
\end{enumerate}

We then obtain 
\begin{equation}
\Phi \left( i,j\right) =\left( 
\begin{array}{ccc}
a & d+e & d-e \\ 
d+e & b+c & f \\ 
d-e & f & b-c
\end{array}
\right) 
\end{equation}
with 
\begin{equation}
|a| > |b| > |c| > |d| > |e| > |f|. 
\end{equation}
\ This hierarchy of force constants is important to keep in mind, given that
the off-diagonal elements of the spring tensors are the most difficult to
obtain from supercell calculations, requiring much bigger supercells than
diagonal elements. There is evidence \cite{avdw:pd3v} that even keeping only
the stretching ($a$) and isotropic bending ($b$) terms of the nearest
neighbor spring tensor can provide vibrational entropies with an accuracy
of about $0.03$ $k_{B}$. If
this observation turns out to be generally applicable, this offers a simple
way to account for vibrational effects in phase diagram calculations. 

\subsubsection{Short-range effective cluster interactions}

If a cluster expansion of the vibrational free energy only requires a small
number of ECI to accurately model the configurational-dependence of the
vibrational free energy, it then becomes practical to determine the values
of these ECI from a small number of very accurate calculations of the
vibrational free energy of a few structures.

The issue of the speed of convergence of the cluster expansion is also
related to the task of devising\ efficient ways to compute vibrational
properties of disordered alloys: The faster the cluster expansion converges,
the easier it is to model a disordered phase (see Appendix \ref{appdisord}).
The calculations of the vibrational entropy change upon disordering has
proven to be a very effective way to assess the importance of lattice
vibrations \cite{althoff:vib,ravelo:vib,avdw:ni3al,avdw:pd3v}, since this
quantity can be straightforwardly used to estimate the effect of lattice
vibrations on transition temperatures with the help of Equation (\ref
{tcshift}).

The central question is thus whether the cluster expansion of the
vibrational free energy converges quickly with respect to the number of ECI.
This is a question distinct from the range of force constants needed to
obtain accurate vibrational properties. The range of ECI needed to represent
the configurational dependence of vibrational free energy may very well
exceed the range of the force constants. Even in simple Born von
K\'{a}rm\'{a}n model systems, there is no direct correspondence between ECI
and force constants, except in special cases (see Section \ref{bpmodel}).
Once relaxations are introduced in the model, then all hope of a simple
correspondence is lost \cite{morgan:local}.

In this context, the question of the existence of a rapidly converging
cluster expansion of vibrational properties has to be answered through
numerical experiments. Simple energy models offer the possibility to test,
at a reasonable computational cost, the speed of convergence of a cluster
expansion. Explicit calculations of a well converged cluster expansion of
vibrational entropy in a Lennard-Jones solid \cite{gdg:vib2} have indicated
that a small number of ECI (9) can provide a good accuracy ($\pm 0.03k_{B}$%
). Other benchmarks of the speed of convergence, based on studies of
disordered alloys \cite{althoff:quasi,morgan:local,morgan:phd,morgan:aruba},
also indicate that concise and accurate cluster expansions are possible.
Experiments that seek to link features of projected phonon DOS to the local
chemical environment of the atoms \cite{fultz:local} suggest that
short-range ECI should be able to successfully model vibrational entropy
differences. One potential source of concern is the difficulty associated
with accounting for the size mismatch effect using a short-range ECI \cite
{morgan:local,avdw:pd3v,morgan:aruba}. In the context of cluster expansions of the
energy, relaxations of the atoms away from their ideal lattice site as a
result of size mismatch are known to introduce both non negligible long
range pair ECI and numerous multiplet ECI. A cluster expansion of the
vibrational free energy is expected to exhibit the same problems. 

All the full phonon DOS ab-initio calculations of vibrational entropies in
alloy systems performed so far have relied on the rapid convergence of the
cluster expansion \cite
{gdg:phd,tepesch:caomgo,avdw:ni3al,ozolins:cuau,avdw:pd3v}. While efforts to
quantify the error introduced by truncating the cluster expansion in
ab-initio calculations have been made \cite{gdg:phd,avdw:ni3al,avdw:pd3v},
the issue of the speed of convergence of the cluster expansion in the
context of vibrational properties clearly merits further study, especially
in light of the importance of the size mismatch effect \cite{avdw:pd3v}.

\section{Experimental techniques}

\label{experimental}The experimental literature on the thermodynamics of
lattice vibrations in alloys relies on mainly three techniques.
In {\em differential calorimetry} measurements, the heat capacity of two
samples in a different state of order is compared over a range of
temperatures. If the upper limit of the range of temperatures is chosen to
be sufficiently low, substitutional exchanges will not occur and the
difference in heat capacity can be assumed to arise solely from vibrational
effects. Integration of the difference in heat capacity (divided by
temperature) then yields a direct measure of the vibrational entropy
differences between the two samples of the range of temperature considered.
This, of course, assumes that the lower temperature bound is sufficiently
low, so that the vibrational entropy of both samples can be assumed to be
zero at that temperature. It also assumes that the electronic contribution
to the heat capacity is negligible. In practice, both assumptions are
typically satisfied. The main problem with this method is that one is
usually interested in vibrational entropy differences at the transition
temperature of the alloy, which is usually above the upper limit of the
temperature range used in the heat capacity measurements. The heat capacity
therefore needs to be extrapolated to high temperature. This constitutes the
main source of inaccuracies in this method. Examples of the use of this
method can be found in \cite
{fultz:ni3alcal,fultz:fe3al,fultz:fe3al2,fultz:cu3au,fultz:ni3v,fultz:co3v}.

A second method is the measurement of phonon dispersion curve through {\em %
inelastic neutron scattering}. For ordered alloys that can be produced in
large single crystals, this method is very powerful. Once the dispersion
curves along special directions in reciprocal space are measured, they can
be used to fit Born-von K\'{a}rm\'{a}n spring models which, in turn, yield
the normal frequencies for any point in the Brillouin zone. With the help of
the standard statistical mechanics techniques described in Section \ref{bvk}%
, this information is sufficient to determine the vibrational entropy.
Examples of applications of this method can be found in \cite
{fultz:fe3al,fultz:fe3al2,fultz:fecr,fultz:ni3alphon,fultz:cu3au}. The
applicability of this method is unfortunately limited by the availability of
large single crystals. The case of disordered alloys presents an even more
fundamental problem: Disordered alloys do not have well defined dispersion
curves and there is no straightforward way to fit the spring constants of a
spring model from the experimental data. This problem is usually addressed
by using the virtual crystal approximation, in which different constituent
atoms are replaced by one ``average'' type of atom (see Appendix \ref
{appdisord}). Unfortunately, this approximation has repeatedly been shown to
have a very limited accuracy for the purpose of measuring vibrational
entropy differences \cite{althoff:vib,ravelo:vib,shaojun:feal,fultz:fe3al2}.
Nevertheless, single crystal phonon dispersion curve measurements for
ordered alloys present a unique opportunity to perform a stringent test of
the accuracy of theoretical models.

A third method is the determination of the phonon density of states from 
{\em incoherent neutron scattering} measurements. In contrast to the
preceding approach, this method can readily be applied to disordered systems
and to compounds for which single crystals are not available \cite
{fultz:ni3alphon,fultz:ni3v,fultz:co3v,fultz:cu3au2,fultz:co3v2}. The main
limitation of this approach is that different atomic species have different
neutron scattering cross-sections. The scattered intensity at each frequency
measures a ``density of states'', where each mode is weighted by the
scattering intensity of the atoms participating in the mode in question.
Thus, one needs some prior information about the vibrational modes in order
to reconstruct the true phonon DOS from the experimental data. In the case
of alloys, there is not a one-to-one correspondence between the measured
data and the vibrational entropy. This problem can be alleviated by choosing
alloy systems where the scattering intensity of each species is similar \cite
{fultz:co3v,fultz:co3v2}.

Other techniques have been used to measure vibrational entropy differences.
Some researchers have used the fact that vibrational entropy and thermal
expansion are directly related, to estimate vibrational entropy differences
from accurate thermal expansion measurements \cite
{mukherjee:fe3al,mukherjee:ni3mn}. The measurement of inelastic nuclear
resonant scattering spectrum has also been used to relate changes in the
phonon DOS to changes in the short-range order of a disordered alloy \cite
{fultz:local}. Finally, relatively noisy estimates of vibrational entropy
differences can be obtained from X-ray Debye-Waller factors or from the
measurement of mean square relative displacement (MSRD) of the atoms
relative to their neighbors through extended electron energy-loss fine
structure (EXELFS) \cite{fultz:ni3alcal}.

\section{Models of lattice vibrations}

\label{modellatvib}While the ability to control the level of approximation
discussed in the previous section is extremely useful, there remains the
problem that, very often, only considering the first few levels in this
hierarchy of approximations already involves substantial computational
requirements. For this reason, models of lattice vibrations that involve
fewer parameters but more physical intuition may provide a practical mean of
including vibrational effects in phase diagram calculations. In this
section, we will present the advantages and weaknesses of each method, in
light of the three fundamental mechanisms described in the Section \ref
{secmecha}.

\subsection{The ``bond proportion'' model}

\label{bpmodel}There have been many attempts (see, for instance, \cite
{Dyson:isot,wojto:latvib,bakker:oddsvib,gdg:vib1}) to find ways to express
the relationship between the vibrational free energy and the dynamical
matrix in a form that illustrates the intuition behind the ``bond
proportion'' mechanism.
In a variety of simple model systems, a convenient exact
expression can be derived for the nearest neighbor ECI in the expansion of
the vibrational free energy in the high temperature limit.
For simple nearest-neighbor spring models with
central forces in linear chains \cite{bakker:oddsvib,matthew:latvib,gdg:vib1},
square lattices \cite{bakker:oddsvib} or
simple cubic lattices \cite{bakker:simple-cubic}, the nearest neighbor ECI is given by
\begin{equation}
V_{1nn}=\frac{d}{8}k_{B}T\ln \left( \frac{k_{AA}k_{BB}}{k_{AB}^{2}}\right) .
\label{eci1nn}
\end{equation}
where $k_{AA}$, $k_{BB}$ and $k_{AB}$ are, respectively, the spring
constants associated with $A-A$, $B-B$ and $A-B$ bonds and $d$ is the
dimensionality of the system. It has been noted, on the basis of numerical
experiments, that the same expression performs well for other lattices \cite
{gdg:vib1}. This success arises from the fact that, as shown in Appendix \ref
{anal1nn},\ Equation (\ref{eci1nn}) is the first order approximation to the
true vibrational entropy change in a large class of systems which satisfies
the following assumptions:

\begin{itemize}
\item  the high temperature limit of the vibrational entropy is appropriate;

\item  the nearest-neighbor force constants can be written as $\Phi \left(
i,j\right) =k_{\sigma _{i}\sigma _{j}}\,\phi \left( i,j\right) $ where $%
k_{\sigma _{i}\sigma _{j}}$ denotes the (scalar) stiffness of the spring
connecting sites $i$ and $j$ with occupations $\sigma _{i}$ and $\sigma _{j}$
while the $\phi \left( i,j\right) $ are dimensionless spring constant
tensors. The $\phi \left( i,j\right) $ are assumed equivalent under a
symmetry operation of the space group of the parent lattice;

\item  all force constants $k_{\sigma _{i}\sigma _{j}}$ are such that 
\begin{equation}
\left| \frac{k_{\sigma _{i}\sigma _{j}}}{\sqrt{k_{\sigma _{i}\sigma
_{i}}k_{\sigma _{j}\sigma _{j}}}}-1\right| \ll 1.  \label{cond1o}
\end{equation}
\end{itemize}

Equation (\ref{eci1nn}) applies to simple harmonic models with nearest
neighbor springs on the fcc, bcc or sc primitive lattices (and,
approximately, on the hcp lattice), as long as the above assumptions are
satisfied. Both stretching and bending terms are allowed in the spring
tensors, as long as their relative magnitude is independent of $k_{\sigma
_{i}\sigma _{j}}$ ({\it e.g.} when the bending terms are always, say, 10\%
of the corresponding stretching term, regardless of the magnitude of the
stretching term). If force constants, other than bending or stretching terms, are important,
the bond proportion model ceases to be valid. This can be seen by the following argument.
The ``bond proportion'' picture requires every bond of a
certain type (for instance, $A-A$ bonds) to have an identical spring
tensor. However, the point symmetry of each bond can be different and
similar chemical bonds in different environment face different
symmetry-induced constraints on their spring tensors \cite{sluiter:cafc}.
The only way to reconcile these observations is use a spring tensor that is
compatible with the highest possible symmetry, ensuring that it is also
compatible with any other environment with a lower symmetry. With the highest
possible symmetry, only two independent terms remain in the spring tensor:
the stretching and bending terms.

Equation (\ref{eci1nn}) embodies the essential intuition behind the effect
of the alloy's state of order on its vibrational free energy.
When one replaces an $A-A$ bond and a $B-B$ bond by
two $A-B$ bonds, the vibrational free energy will decrease only if the
stiffness of $A-B$ bonds, $k_{AB}$, exceeds the geometrical average
stiffness of the bonds between identical species $\sqrt{k_{AA}k_{BB}}$. This
observation allows the determination of the expected effect of vibrations on
the shape of the phase diagram by simple arguments. The link between the
nearest neighbor ECI of the expansion of the vibrational entropy can be
summarized by the expression \cite{gdg:vib2}: 
\begin{equation}
\frac{T_{c}^{\text{config+vib}}}{T_{c}^{\text{config}}}=\frac{1}{1\mp \alpha
V_{1nn}/k_{B}T} 
\end{equation}
where the ``$-$'' and ``$+$'' correspond to ordering and segregating
systems, respectively, and where $\alpha $ is a dimensionless parameter that
only depends on the lattice type and the ordering tendency of the system
(for instance, for fcc, $\alpha =1.7$ in ordering systems and $\alpha =9.8$
in segregating systems, while for bcc, $\alpha =6.5$ in both cases).

It is straightforward to include vibrational effects in phase diagram
calculations using the ``bond proportion'' model. All that is needed is an
estimate of the stiffness of $A-A$, $B-B$ and $A-B$ bonds, which could come,
for instance, from supercell calculations of the nearest neighbor force
constants in a few simple structures or from the bulk moduli of the pure
elements and one ordered compound. The nearest neighbor ECI then obtained
can be simply added to the cluster expansion of the energy.

While Equation (\ref{eci1nn}) is useful to estimate the importance of the
``bond proportion'' mechanism in a given system, one can avoid some of the
approximations involved in deriving Equation (\ref{eci1nn}) at the expense
of only a modest amount of additional effort. One can find the
exact phonon DOS of the nearest neighbor Born-von K\'{a}rm\'{a}n model for a
variety of configurations of the alloy, which allows a more accurate cluster
expansion of the vibrational energy to be derived. In this fashion, the
condition specified in Equation (\ref{cond1o}) is no longer needed and the
vibrational entropy can be calculated at any temperature.

It is important to keep in mind that two important assumptions are made when
invoking the ``bond proportion'' mechanism. First, vibrational entropies are
solely determined by the nearest neighbor force constants. There is
theoretical evidence that nearest neighbor spring models can predict
vibrational entropy differences with an accuracy of about $0.02k_{B}$ in
metallic \cite{avdw:ni3al,avdw:pd3v} and semiconductor \cite{gdg:phd}
systems. Given that configurational entropy differences are typically of the
order of $0.2k_{B}$, this precision should be sufficient for practical phase
diagram calculations.

The second assumption is that each type of chemical bond is assumed to have
an intrinsic stiffness that is independent of its environment.
First-principles calculations on the Li-Al \cite{sluiter:cafc} and on the
Pd-V system \cite{avdw:pd3v} unfortunately indicate that the stiffness of a
chemical bond does change substantially as a function of its environment.
This problem is serious, as it considerably limits the applicability of the
``bond proportion'' model. These changes of the intrinsic stiffness of the
bonds as a function of their environment are precisely the focus of the two
other suggested sources of vibrational entropy changes. In summary, while
the ``bond proportion'' model gives an elegant description of one of the
mechanisms suggested to be at the origin of vibrational entropy differences,
it completely ignores the two other mechanisms, namely, the volume and size
mismatch effects.

\subsection{The Debye model}

Perhaps the most widespread approximation to the phonon DOS $g(\nu )$ is the
Debye model (see, for instance, \cite{grimvall:thermophys,ashmer:sol}), where the phonon problem is solved
in the acoustic limit. In this case, the phonon DOS is approximated by: 
\begin{equation}
g\left( \nu \right) =\left\{ 
\begin{array}{c}
\frac{9\nu ^{2}}{\nu _{D}^{3}}\text{ if }\nu \leq \nu _{D} \\ 
0\text{ if }\nu >\nu _{D}
\end{array}
\right. 
\end{equation}
where $\nu _{D}=\frac{k_{B}\Theta _{D}}{h}$ and $\Theta _{D}$ is the Debye
temperature, given by: 
\begin{equation}
\Theta _{D}=\frac{h}{k_{B}}\left( \frac{3N}{4\pi V}\right) ^{1/3}C_{D} 
\end{equation}
where $C_{D}$ is the Debye sound velocity, defined by 
\begin{equation}
\frac{3}{C_{D}^{3}}=\left\langle \sum_{\lambda =1}^{3}\frac{1}{C_{\lambda
}^{3}}\right\rangle 
\end{equation}
where the right-hand side is the directional average of a function of the
three sound velocities $C_{\lambda }$ \cite{grimvall:thermophys}.\ The free
energy of a Debye solid is given by: 
\begin{eqnarray*}
\frac{F}{N} &=&\frac{E^{\ast }}{N}+\frac{9}{8}k_{B}\Theta _{D}-k_{B}T\left(
D\left( \frac{\Theta _{D}}{T}\right) -3\ln \left( 1-\exp \left( -\frac{%
\Theta _{D}}{T}\right) \right) \right) \\
&\approx &\frac{E^{\ast }}{N}+3k_{B}T\ln \left( \frac{k\Theta _{D}}{h}%
\right) \text{ in the high temperature limit.}
\end{eqnarray*}
where the Debye function $D\left( u\right) $ is given by 
\begin{equation}
D\left( u\right) =3u^{3}\int_{0}^{u}\frac{x^{4}e^{x}}{\left( e^{x}-1\right)
^{2}}dx 
\end{equation}

Since the Debye sound velocity $C_{D}$ is a complicated function of all
elastic constants of the material, an approximation to the Debye temperature
that only involves the bulk modulus proves extremely useful. Such an
approximation was derived by Moruzzi, Janak and Schwarz (MJS) \cite
{moruzzi:debyeg} for cubic materials\footnote{%
although it has been used for materials with a lower symmetry \cite
{asta:cdmg,sanchez:agcu}}: 
\begin{equation}
\Theta _{D}=0.617\left( \frac{3}{4\pi }\right) ^{1/3}\frac{h}{k_{B}}\left( 
\frac{\Omega ^{1/3}B}{\overline{M}}\right) ^{1/2} 
\end{equation}
where $\Omega $ is the average atomic volume $B$ is the bulk modulus and $%
\overline{M}$ is the concentration weighted arithmetic mean of the atomic
masses. As noted in \cite{gdg:vib1}, in the high temperature limit, the MJS
model does not exhibit the required property that the masses have no effect on the
vibrational free energy of formation, although using a geometric average of
the masses \cite{gdg:vib2} fixes this problem.

The quasiharmonic approximation can be used, within Debye theory, to account
for mild anharmonicity. In the so-called Debye-Gr\"{u}neisen approximation,
the volume-dependence of the phonon DOS is modeled by a single Gr\"{u}neisen
parameter and the effect of volume can be summarized by simply making the
Debye temperature volume-dependent: 
\begin{equation}
\Theta _{D}=\Theta _{D,0}\left( \frac{V_{0}}{V}\right) ^{\gamma }. 
\end{equation}
where $\Theta _{D,0}$ is the Debye temperature at volume $V_{0}$ and $\gamma 
$ is the gr\"{u}neisen parameter.

Despite of its inaccurate description of the true phonon DOS at high
frequencies, the Debye and Debye-Gr\"{u}neisen models are quite successful
at modeling the changes in vibrational properties of a given compound as a
function of temperature. For instance, the thermal properties of pure metals 
\cite{moruzzi:debyeg} calculated in MJS approximation are surprisingly
accurate. The reason for this success is that most thermodynamic quantities
({\it e.g.} free energy, entropy, heat capacity, etc.) exhibit
their most dramatic variations at low temperature, where the low frequency
phonon modes that are correctly described by the Debye model have a dominant
effect. In the high temperature regime, thermodynamic quantities are
determined by the classical equipartition theorem, and any harmonic model
gives the correct behavior.

Debye-like models are expected to perform well in systems where the
differences in vibrational free energy between compounds can be explained by
uniform shifts of the phonon DOS, such as when the volume effect operates
alone. Such a behavior has been observed in Embedded Atom calculations on
the Ni--Al system \cite{ackland:vol,althoff:vib,ravelo:vib} but in no other
systems so far. The MJS approximation has been used to include vibrational
effects in phase diagram calculations and has resulted in an improved
agreement with experimental results \cite
{asta:cdmg,sanchez:agcu,Colinet:AuNi}.

However, as shown in \cite{gdg:vib2}, the Debye approximation and its
successors can have significant shortcomings when used to calculate phase
diagrams. A significant part of the vibrational free energy differences
between different compounds arises from changes in the high frequency
portion of the phonon DOS, which Debye-like models describe incorrectly. In
some cases, the MJS approximation can even lead to an incorrect prediction
of the sign of the vibrational entropy difference \cite
{avdw:pd3v,ozolins:cuau}.

In summary, the Debye model and its derivatives capture the essential
physics behind only one of the advocated mechanisms responsible for the
configurational dependence of vibrational free energy: the volume effect.
Approximations based on the Debye model, however, fail to account for the
possibility that the state of order also has a direct impact ({\it i.e.}
not through the volume) on the shape of
the phonon DOS (as predicted, for instance, by the ``bond proportion''
model).

\subsection{The Einstein model}

The perfect complement to the Debye model is the Einstein model (see, for instance,
\cite{ashmer:sol,mcquarrie:stattherm}), which
focuses on the high frequency region of the phonon DOS, instead of its low
frequency region. The Einstein model assumes that a crystalline solid can be modelled
by a collection of $3N$ independent harmonic oscillators (3 per atom)
sharing a common frequency.
This frequency can, for instance, be determined by computing the natural frequency of
oscillation of one atom when all others are frozen in place.
This approach, known as the local harmonic
model \cite{Srolovitz:local-harmonic1,Sutton:local-harmonic}, has proven
especially useful to calculate vibrational entropies associated with defects 
\cite{Srolovitz:vacancy,Srolovitz:local-harmonic2,Sutton:local-harmonic}.
The Einstein model can also be combined with a Debye model to better fit
experimental calorimetry data \cite{fultz:fe3al} or thermal expansion data 
\cite{mukherjee:fe3al}. 

The local harmonic model is of little use whenever the system of interest
exhibits translational symmetry, because the calculations required
to determine the unknown parameters of an Einstein model
from first-principles directly provide force constants. The latter could
be used to obtain a more precise description of the DOS
rather than the single-value DOS characterizing the Einstein model.

The Einstein model is nevertheless extremely useful for conceptual purposes,
as we will now illustrate. As shown in Appendix \ref{einstein}, the
vibrational free energy of a system is bounded by above and by below by the
free energy of two Einstein-like systems: 
\begin{equation}
\frac{k_{B}T}{2}\ln \left( \frac{h}{k_{B}T}\prod_{i}M_{i}\left( \left( \Phi
^{-1}\right) _{ii}\right) ^{-1}\right) \leq F_{\text{vib}}\leq \frac{k_{B}T}{%
2}\ln \left( \frac{h}{k_{B}T}\prod_{i}M_{i}\Phi _{ii}\right) 
\end{equation}
While the upper bound is obtained from the usual local harmonic model, where
surrounding atoms do not relax, the lower bound is obtained when the
surrounding atoms are allowed to relax freely. Another way to interpret
these bounds is that at one extreme, each atom sees the others as having an
infinite mass, while at the other extreme, each atom sees the other atoms as
being massless. This result supports the view that vibrational free energy
can be meaningfully considered as a measure of the average stiffness of each
atom's local environment.

A more rigorous way of defining the contribution of an atom to the total
vibrational free energy is the use of the projected DOS (see, for instance,
\cite{morgan:local}). This approach does not in any way simplify the calculation of
vibrational properties, because the full phonon DOS is needed as an input,
but it is a useful way to interpret the experimentally measured or
calculated phonon DOS. To obtain the contribution of atom $i$ to the DOS,
the idea is to weight each normal mode by the magnitude of the vibration of
atom $i$: 
\begin{equation}
g_{i}\left( \nu \right) =\frac{1}{N}\sum_{j}\left| e_{j}\left( i\right)
\right| ^{2}\delta \left( \nu -\nu _{i}\right) 
\end{equation}
where $e_{j}$ is the eigenvector (normalized to unit length) associated with
the mode of frequency $\nu _{j}$. Since extensive thermodynamic properties
are linear in the DOS, atom-specific local thermodynamic properties can be
readily defined from the projected DOS. Note that, by construction, all the
projected DOS sum up to the true phonon DOS and thus, all the local
extensive thermodynamic quantities sum up to the corresponding total
quantity. 

\subsection{The ``bond stiffness vs. bond length'' approach}

In {\it ab-initio} calculations, most of the computational burden comes from
the calculation of the force constant tensors. It would thus be extremely
helpful if the force constants determined in one structure could be used to
predict force constants in another structure. From the failures of the
``bond proportion'' model, however, we know that forces constants obtained
from one structure are not directly transferable to another structure \cite
{sluiter:cafc,avdw:pd3v}

Nevertheless, a simple modification of the transferable force constant
approach yields substantial improvements in precision. First-principles
calculation the Pd-V \cite{avdw:pd3v} system revealed that most of the
variation in the stiffness of a given chemical bond across different
structures can be explained by changes in bond length.\footnote{%
The results obtained in the Li-Al system \cite{sluiter:cafc} also suggest
than length is a good predictor of stiffness, although these authors
did not investigate this matter further.} This suggests that the transferable quantity to
consider is a ``bond stiffness vs. bond length'' relationship. As a
first approximation, a linear relationship can be used 
\begin{eqnarray*}
a\left( l\right) &=&a_{0}+a_{1}\left( l-l_{0}\right) \\
b\left( l\right) &=&b_{0}+b_{1}\left( l-l_{0}\right)
\end{eqnarray*}
where $a$ and $b$ denote the stretching and isotropic bending terms,
respectively and where $a_{0}$ and $b_{0}$ describe the stiffness of the
bond at its ideal length $l_{0}$ while $a_{1}$ $\ $and $b_{1}$ are analogous
to bond-specific Gr\"{u}neisen parameters. The other parameters of the
spring tensor are unlikely to follow such simple relationships because
they may be required to vanish according to the local symmetry
of the bond, independently of its length (this is discussed in more details
in Appendix \ref{anal1nn}).

This approximation was shown to be successful in the Pd-V system \cite
{avdw:pd3v}. Figure \ref{fitpp} illustrates the ability of this simple model
to predict bond stiffness in different structures. A similar analysis
performed with the data on the Ni-Al system from Reference \cite{avdw:ni3al} 
is shown in Fig. \ref{nialfkall}. Table \ref{svslpred} compares the
predictions obtained from the ``bond stiffness vs. bond length'' model with more
accurate calculations.

There are numerous advantages to this approach. From a conceptual point of
view, this model presents a concise way to represent all three mechanism
suggested to be the source of vibrational entropy differences. The ``bond
proportion'' mechanism is the particular case obtained when little changes
in bond length occur. The volume effect results from expanding all bonds by
the same factor. The size mismatch effect (or the ``stiff sphere'' picture)
is also modeled since the local change in stiffness resulting from locally
compressed or expanded regions are explicitly taken into account. A
straightforward way to represent the source of vibrational changes is to
overlap the stiffness vs. length relationship and the changes in
average length and stiffness in different states of order, as shown in
Fig. \ref{shiftstiff}.

A second advantage of this method is computational efficiency. The unknown
parameters of the model can be determined by a small number of supercell or
linear response calculations. After that, the knowledge of the relaxed
geometry of a structure is sufficient to determine the stiffness of all
chemical bonds. Finding the vibrational entropy of the structure then just
reduces to a computationally inexpensive Born-von K\'{a}rm\'{a}n phonon
problem. It is important to note that
the knowledge of the relaxed geometries of a set of structures is a natural
by-product of first-principles calculations of structural energies, which
are needed to construct the cluster expansion of the energy in phase
diagram calculations, whether vibrational effects are included or not. Since
computational requirements do not grow rapidly with the number of structures
considered, this opens the way for a much more accurate representation of
the configurational-dependence of the vibrational free energy.

A third advantage of transferable bond stiffness vs. bond length relationships is
that they contain all the information needed to account for thermal
expansion as well, within the quasi-harmonic approximation. The slopes of
the stiffness vs. length relationships for each chemical bonds explicitly
defines the changes in phonon frequencies as volume changes. Since the bulk
modulus of each structure is also a by-product of structural energy
calculations, all the ingredients needed for a quasi-harmonic treatment of
thermal expansion are available.

\section{Conclusion}

Lattice vibrations can have a significant impact on phase transition temperatures,
short-range order, solubility limits, and the sequence in which phases
appear as a function of temperature.
The standard framework of alloy theory can be straightforwardly extended to
account for lattice vibrations using the concept of coarse graining of the
partition function. Once the degrees of freedom associated with lattice
vibrations are integrated out, one is left with a standard Ising model,
where the energy of each spin configuration is replaced by its vibrational
free energy. The efficient evaluation of the vibrational free energy of each
configuration is the main problem limiting the inclusion of lattice
vibrations in phase diagram calculations. A number of investigations have
sought to assess the importance of vibrational effects on phase stability,
in order to ensure that the efforts involved in computing vibrational
properties are justified. The conclusion of the most reliable of these
studies is that vibrational entropy differences are typically on the order
of $0.1$ $k_{B}$ to $0.2$ $k_{B}$, which is comparable to the magnitude of
configurational entropy differences (at most $0.69$ $k_{B}$ in binary alloys),
thereby indicating that vibrations have a nonnegligible impact.

The calculation of the vibrational free energy of a particular configuration of the
alloy reduces to the well known phonon problem in crystals. While the
standard harmonic treatment of this problem lacks the ability to model
thermal expansion, which can have a significant impact on thermodynamic
properties in alloys, this limitation is easily overcome with the help of
the quasiharmonic model. An exact solution to the phonon problem for all
possible configurations requires excessive computing power. However, the
tradeoff between accuracy and computational requirements can be controlled
in two ways, namely through the selection of the range of force constants in
the Born-von K\'{a}rm\'{a}n model, and through a choice of the number of ECI
used to expand the configuration dependence of the vibrational free energy.
While there is evidence that the range of force constants can be kept
very small (first nearest neighbor springs), the configurational dependence
of the vibrational free energy is too complex to permit a drastic reduction
in the number of ECI.

Given the substantial computing power required to undertake lattice dynamics calculations,
many attempts have been made to devise simpler models.
For many years, the MJS approximation appeared to be a very promising way to
include vibrational effects in phase diagram calculations, because it systematically
improved the agreement between first-principles calculations and experimental
measurements. This success may have been the result of two fortunate circumstances
(i) First-principles phase diagram calculations typically overestimate transitions
temperature and (ii) the MJS approximation nearly always yields a downward correction
to the transition temperature. As the accuracy of phase diagram calculations improved
through the use of longer-range cluster expansions
\cite{tepesch:caomgo,ozolins:noble,vanderven:licoo2},
the systematic bias in the calculated transitions temperature substantially decreased.
Simultaneously, more sophisticated models of lattice vibrations indicated that lattice
vibrations do not always results in a reduction in the transition temperatures
\cite{gdg:phd,gdg:vib2,avdw:pd3v}. The net effect of these two trends is that, although
the accuracy of first-principles calculations has increased over the years, obtaining
improved agreement with experiment is now a much more stringent test. As a result,
perfectly valid and accurate calculations of
vibrational effects sometimes reduce the agreement with experiments
\cite{tepesch:caomgo,ozolins:cuau}. Hence, before one can unambiguously assess
the importance of lattice vibrations through a full phase diagram calculation,
all potential sources of error have to be carefully controlled, such as the
precision of the energy model used and more importantly, the accuracy of the
cluster expansion. To date, the most convincing evidence that taking lattice
vibrations into account significantly improves agreement with experimental
results comes from calculations of the lattice dynamics associated with a
specific atomic configuration ({\it e.g.} a given compound or an isolated
point defect) \cite{ozolins:alsc,wolverton:cual,quong:thexp}. In these settings, most
sources of errors are under control and definite answers can be given.
 
Although the availability of more accurate computational tools has revealed that
the trends in vibrational entropy differences between phases is far more complex that anticipated
ten years ago (\cite{gdg:vib2,althoff:vib,avdw:pd3v,morgan:aruba,sluiter:cafc,wolverton:cual}),
a simple picture of the mechanisms at work is now emerging.
All the known sources of vibrational entropy differences
can be conveniently summarized by the ``bond stiffness vs.
bond length'' model \cite{avdw:pd3v,morgan:aruba}.
In this picture, each type of chemical bond is characterized by a
length-dependent spring constant. Changes in vibrational entropy can
originate from both changes in the proportion of each chemical bond and
changes in their lengths as a result of local and global relaxations. This
model not only provides an intuitive understanding of lattice vibrations in
alloys, but also a practical way of including their effects in phase diagram
calculations. This stiffness vs. length relationship of each type of
chemical bond can be inferred from a small number of lattice dynamics
calculations. The vibrational properties of any configuration can then be
obtained at a very low computational cost from the knowledge of the
equilibrium geometry of this configuration, an information that is already a
natural by-product of any phase diagram calculation.

Future investigations of the effect of lattice vibrations on phase stability
should head towards three main directions.
\begin{enumerate}
\item While reporting error bars is an important part of any experimentalists' work,
theorists should devote significantly more effort to quantifying the uncertainties 
of their calculations. This would make it possible to clearly identify situations where 
the improved agreement with experimental results following the inclusion of vibrational 
effects is truly significant or merely the result of fortunate coincidences. It is 
admittedly difficult to quantify the errors introduced by
the energy model (such as the LDA), but standard statistical techniques can clearly 
be used to quantify any error due to fitting the {\it ab initio} data with 
a simple model.
\item Given the difficulty of extracting vibrational entropies from experimental data, 
theorists should undertake the computation of quantities that {\em can} be directly 
measured. For instance, a Born-von K\'arm\'an model directly
enables the simulation of incoherent neutron scattering data, while the inverse procedure 
is a highly non unique operation. The calculation of thermal expansion coefficients 
would also be a very sensitive test.
\item There have so far been very few accurate phase diagram calculations that
include the effect of lattice vibrations. The main limitation remains the
determination of a cluster expansion that accurately models the
configurational dependence of vibrational free energy. The ``bond length vs.
bond stiffness'' model should prove to be an extremely useful tool in
achieving this goal. Although this approximation has been very successful in
all systems to which it has been applied, the confirmation of its validity
in a wider range of systems is crucial. It would also be interesting to devise
a hierachy of increasingly accurate approximations that would include the ``bond length vs.
bond stiffness'' model as a particular case.
\end{enumerate}

\section*{Acknowledgements}

This work was supported by the U.S. Department of Energy, Office of
Basic Energy Sciences, under contract no. DE-F502-96ER 45571.
Gerbrand Ceder acknowledges support of Union Mini\`ere through
a Faculty Development Chair. 
Axel van de Walle acknowledges support of the National Science Foundation
under program DMR-0080766 during his stay at Northwestern
University.

\appendix

\section{The absence of mass effects in the high-tempe{\-}ra{\-}ture limit}

\label{masscancel} This appendix shows that 
the vibrational entropy of formation is independent of
the atomic masses in the high temperature limit, as
several authors \cite{grimvall:nomass,gdg:vib2} have noted.
In the high-temperature limit, the vibrational entropy is
determined by the product of the frequencies of all normal modes of
vibrations $\nu _{m}$, which can be related to the eigenvalues $\lambda _{m}$
of the $3N\times 3N$ dynamical matrix $D$ of the system (up to a constant):%
\footnote{%
To simplify the exposition and avoid the problem that the dynamical matrix
has three zero eigenvalues associated with the possibility of a rigid
translation of the system, we assume that some of the atoms of the system
are attached to a fixed point of reference by a weak spring. In the
thermodynamic limit, this assumption becomes inconsequential.} 
\begin{eqnarray*}
\sum_{m}\ln \left( \nu _{m}\right) &=&\ln \left( \prod_{m}\nu _{m}\right) =%
\frac{1}{2}\ln \left( \prod_{m}\lambda _{m}\right) +\text{const} \\
&=&\frac{1}{2N}\ln \left( \det D\right) +\text{const}
\end{eqnarray*}
Using the properties of determinants, we can write: 
\begin{eqnarray*}
\frac{1}{2N}\ln \left( \det D\right) &=&\frac{1}{2N}\ln \left( \det \left(
M^{1/2}\Phi M^{1/2}\right) \right) \\
&=&\frac{1}{2N}\ln \left( \det \left( \Phi \right) \det \left( M\right)
\right) =\frac{1}{2N}\ln \left( \det \left( \Phi \right)
\prod_{i}M_{i}^{3}\right) \\
&=&\frac{1}{2N}\ln \left( \det \left( \Phi \right) \right) +\frac{3}{2N}%
\sum_{i}\ln M_{i}
\end{eqnarray*}
where $M$ is the $3N\times 3N$ diagonal matrix of all the $N$ atomic masses
of the system (each repeated three times) while $\Phi $ is $3N\times 3N$ the
matrix obtained by regrouping all the $3\times 3$ force constant tensors $%
\Phi \left( i,j\right) $ in a single matrix (analogously to Equation (\ref
{bigdynmat})). Now consider the change in the value of $\sum_{m}\ln \left(
\nu _{m}\right) $ when an $N_{A}$ atoms of type $A$ and $N_{B}$ atoms of
type $B$ are combined to form an alloy. Let the subscripts $AB$, $A$ and $B$
respectively denote the properties of an $A_{\left( N_{A}/N\right)
}B_{\left( N_{B}/N\right) }$ alloy, a pure crystal of element $A$ and a pure
crystal of element $B$. 
\begin{eqnarray*}
\Delta \left( \sum_{m}\ln \left( \nu _{m}\right) \right) &=&\sum_{m}\ln
\left( \nu _{m}^{AB}\right) -x_{A}\sum_{m}\ln \left( \nu _{m}^{A}\right)
-x_{B}\sum_{m}\ln \left( \nu _{m}^{B}\right) \\
&=&\frac{1}{2}\ln \left( \det \left( \Phi ^{AB}\right) \right) +\frac{3}{2}%
N_{A}\ln M_{A}+\frac{3}{2}N_{B}\ln M_{B} \\
&&-\frac{1}{2}\ln \left( \det \left( \Phi ^{A}\right) \right) -\frac{3}{2}%
N_{A}\ln M_{A}-\frac{1}{2}\ln \left( \det \left( \Phi ^{B}\right) \right) -%
\frac{3}{2}N_{B}\ln M_{B} \\
&=&.\frac{1}{2}\ln \left( \det \left( \Phi ^{AB}\right) \right) -\frac{1}{2}%
\ln \left( \det \left( \Phi ^{A}\right) \right) -\frac{1}{2}\ln \left( \det
\left( \Phi ^{B}\right) \right)
\end{eqnarray*}
All the terms involving masses exactly cancel one another.

\section{A simple model of anharmonicity}

\label{anhapp} The material presented in this appendix combines
standard results regarding the Gr\"uneisen framework
that can be found, for instance, in \cite{grimvall:thermophys}.

Two assumptions are made. First, the elastic energy of the
motionless lattice is assumed quadratic in volume: 
\begin{equation}
E^{\ast }\left( V\right) =\frac{B}{2V_{0}}\left( \Delta V\right) ^{2} 
\end{equation}
where $B$ is the bulk modulus, $V_{0}$ the equilibrium volume at $0K$
(ignoring zero-point motion) and $\Delta V=V-V_{0}$. Second, the high
temperature limit of the vibrational free energy is used: 
\begin{equation}
F_{vib}\left( T,V\right) =k_{B}T\sum_{i}\ln \left( \frac{h\nu _{i}}{k_{B}T}%
\right) 
\end{equation}
In this approximation, the volume-dependence of $F_{vib}$ takes on a
particularly simple form: 
\begin{equation}
\frac{\partial F_{vib}\left( T,V\right) }{\partial V}=\frac{3Nk_{B}T%
\overline{\gamma }}{V} 
\end{equation}
where 
\begin{equation}
\overline{\gamma }=\frac{1}{3N}\sum_{i=1}^{3N}\frac{V}{\nu _{i}}\frac{%
\partial \nu _{i}}{\partial V} 
\end{equation}
is an average Gr\"{u}neisen parameter. In the high-temperature limit, an
average Gr\"{u}nei{\-}sen parameter can easily be defined, because the
population of the phonon modes is no longer temperature-dependent, and any
change in entropy can be unambiguously attributed to shifts in phonon
frequencies. At lower temperatures, the changes in phonon population would
need to be accounted for as well.

If we assume that the volume-dependence of the vibrational free energy is
linear in volume, we have: 
\begin{eqnarray*}
F\left( T,V\right) &=&E^{\ast }\left( V\right) +F_{vib}\left( T,V\right) \\
&=&E^{\ast }\left( V\right) +F_{vib}\left( T,V_{0}\right) +\left. \frac{%
\partial F_{vib}}{\partial V}\right| _{V=V_{0}}\Delta V \\
&=&\frac{B}{2V_{0}}\left( \Delta V\right) ^{2}+F_{vib}\left( T,V_{0}\right) +%
\frac{3Nk_{B}T\overline{\gamma }}{V_{0}}\Delta V.
\end{eqnarray*}
Minimizing this expression with respect to $\Delta V$ yields: 
\begin{equation}
\frac{\Delta V}{N}=\frac{3k_{B}T\overline{\gamma }}{B} 
\end{equation}
where $\frac{3k_{B}T\gamma }{B}$ is the coefficient of volumetric thermal
expansion. The resulting temperature dependence of the free energy (for one
given configuration ${\bf\sigma}$) is given by 
\begin{equation}
\frac{F\left( T\right) }{N}=\frac{F\left( T,V_{0}\right) }{N}-\frac{\left(
3k_{B}T\overline{\gamma }\right) ^{2}}{2B\left( V_{0}/N\right) }. 
\end{equation}
It is interesting to note that half of the vibrational free energy decrease
due to thermal expansion is canceled by the energy increase of the
motionless lattice. Hence, vibrational entropy differences originating from
differences in thermal expansion between phases have, relative to other
sources of vibrational entropy changes, half the effect on phase stability.

\section{Modeling the disordered state}

\label{appdisord}Although in phase diagram calculations, the use of the
cluster expansion bypasses the problem of directly calculating the
vibrational entropy of a disordered phase, there are cases where it is of
interest to directly calculate the vibrational properties of the disordered
state. For instance, in studies that seek to assess the importance of
lattice vibrations \cite{althoff:vib,ravelo:vib,avdw:ni3al,ozolins:cuau},
it is instructive to compute the vibrational entropy change upon disordering
an alloy, since this quantity can be straightforwardly used to estimate the
effect of lattice vibrations on transition temperatures with the help of
Equation (\ref{tcshift}). Here are the most common methods used to model the
disordered state.

Perhaps the most obvious and brute force approach to modeling the disordered
state is the use of a large supercell where the occupation of each site is
chosen at random. This approach was chosen in all EAM calculations \cite
{ackland:vol,althoff:vib,ravelo:vib,morgan:local} as well as in other
investigations \cite{shaojun:feal}. Unfortunately, it is generally not
feasible in the case of {\it ab initio} calculations.

The virtual crystal approximation (VCA) consists of replacing each atom in a
disordered alloy by an ``average'' atom whose properties are determined by a
concentration weighted average of the properties of the constituents. In the
limit where the chemical species have nearly identical properties, this
approximation is justified. This model has been commonly used to interpret
neutron scattering measurements of phonon dispersion curves in the case of
disordered alloys \cite{fultz:ni3alphon,fultz:cu3au}. It has also been used
in a some theoretical investigations \cite{cleri:cu3au,persson:rew}.
However, the virtual crystal approximation has been repeatedly shown to be
insufficiently accurate for the purpose of calculating differences in
vibrational entropies between distinct compounds \cite
{althoff:vib,ravelo:vib,shaojun:feal,fultz:fe3al2}. Its weaknesses are
numerous: It is unable to model ``bond proportion'' effects, volume effects
and local relaxations. It also fails to give a mass-independent high
temperature limit.

Special Quasirandom Structures (SQS) \cite{zunger:sqs} combine the idea of
cluster expansion with the use of supercells. SQS are the periodic
structures that best approximate the disordered state in a unit cell of a
given size. The quality of a SQS is described by the number of its
correlations that match the ones of the true disordered state. There is thus
a direct connection between cluster expansions and SQS: If there exists a
truncated cluster expansion that is able to predict properties.of the
disordered state there exists an SQS that provides an equally accurate
description of the disordered state.

SQS have been very successfully used to obtain electronic and thermodynamic
properties of disordered materials (see, for example, \cite{zunger:sqstest}%
).The accuracy of the SQS approach in the context of phonon calculations has
been benchmarked using embedded atoms potentials which allow for the
comparison with the ``exact'' vibrational entropy of the disordered state
with a large supercell \cite{morgan:local}. It has been found that, for the
purpose of calculating vibrational properties, an SQS having only 8 atoms in
its unit cell already provide a good approximation of the disordered state
in the case of an fcc alloy at concentration $3/4$. While the performance of
this small SQS is remarkable in a model system where local relaxations are
disallowed, it tends to degrade somewhat when relaxations are allowed to
take place. This effect can naturally be explained by the fact that
relaxations are known to introduce important multibody terms in the cluster
expansion, which translates into the requirement that the SQS must correctly
reproduce the corresponding multibody correlations.

The success of small SQS opened the way for the use of more accurate energy
models to calculate vibrational properties of disordered alloys. SQS have
been applied to the {\it ab-initio} calculation of vibrational entropy in
disordered Ni$_{3}$Al and Pd$_{3}$V alloys \cite{avdw:ni3al,avdw:pd3v}

\section{The Einstein model}

\label{einstein}In the Einstein model of a solid, the free energy, in the
high temperature limit, is given by 
\begin{eqnarray*}
F &=&k_{B}T\ln \left( \frac{h}{k_{B}T}\prod_{i}\nu _{i}\right) \\
&=&\frac{k_{B}T}{2}\ln \left( \frac{h}{k_{B}T}\det D\right) \\
&=&\frac{k_{B}T}{2}\ln \left( \frac{h}{k_{B}T}\det \left( M^{1/2}\Phi
M^{1/2}\right) \right) \\
&=&\frac{k_{B}T}{2}\ln \left( \frac{h}{k_{B}T}\det M\det \Phi \right)
\end{eqnarray*}
where $D$ and $\Phi $ are, respectively, the $3N\times 3N$ dynamical matrix
and force constant matrix of the system while $M$ is the matrix of the
masses: 
\begin{equation}
M_{ij}=\delta _{ij}M_{j}. 
\end{equation}
It can be shown \cite{avdw:master} that for any positive definite matrix $%
\Phi $%
\begin{equation}
\det \Phi \leq \prod_{i}\Phi _{ii}, 
\end{equation}
implying that 
\begin{equation}
F\leq \frac{k_{B}T}{2}\ln \left( \frac{h}{k_{B}T}\prod_{i}M_{i}\Phi
_{ii}\right) 
\end{equation}
where the right-hand side expression is nothing but the free energy of the
system in the Einstein approximation. A lower bound can be obtained by a
similar technique, by using the inverse of the force constant matrix 
\begin{equation}
\det \Phi \geq \left( \prod_{i}\left( \Phi ^{-1}\right) _{ii}\right) ^{-1}. 
\end{equation}
The interpretation of the inverse $\Phi $ is simple: It is the matrix that
maps forces $F$ exerted on the atoms to the resulting displacements $u$ of
the atoms. 
\begin{equation}
u=\Phi ^{-1}F 
\end{equation}
While $\Phi _{ii}$ is related to the oscillation frequency of a single atom
when all other atoms are held in place, $\left( \left( \Phi ^{-1}\right)
_{ii}\right) ^{-1}$ is related to the oscillation frequency of an atom $i$
when all surrounding atoms are allowed to relax so that the force exerted on
them remains zero as atom $i$ moves. Atom $i$ has mass $M_{i}$ while all
other atoms are considered massless and relax instantaneously. Atoms located
infinitely far away from atom $i$ are held in place with an infinitesimal
force.

In conclusion, the free energy of a system is bounded by above and by below
by the free energy of two Einstein-like systems: 
\begin{equation}
\frac{k_{B}T}{2}\ln \left( \frac{h}{k_{B}T}\prod_{i}M_{i}\left( \left( \Phi
^{-1}\right) _{ii}\right) ^{-1}\right) \leq F\leq \frac{k_{B}T}{2}\ln \left( 
\frac{h}{k_{B}T}\prod_{i}M_{i}\Phi _{ii}\right) 
\end{equation}

\section{Derivation of the ``bond proportion'' model}

\label{anal1nn}This appendix generalizes the results found in
\cite{bakker:oddsvib,matthew:latvib,gdg:vib1,bakker:oddsvib,bakker:simple-cubic}
in order to handle more general lattice types.
We show that, in an important class of systems,
the bond proportion model is in fact the first order approximation to the
true change in vibrational entropy induced by a change in the proportion of
the different types of chemical bonds.

The alloy system is assumed to satisfy the following conditions:

\begin{itemize}
\item  the high temperature limit is appropriate;

\item  the nearest-neighbor force constants can be written as $\Phi \left(
i,j\right) =k_{\sigma _{i}\sigma _{j}}\,\phi \left( i,j\right) $ where $%
k_{\sigma _{i}\sigma _{j}}$ denotes the (scalar) stiffness of the spring
connecting sites $i$ and $j$ with occupations $\sigma _{i}$ and $\sigma _{j}$
while the $\phi \left( i,j\right) $ are dimensionless spring constant
tensors. The $\phi \left( i,j\right) $ are assumed equivalent under a
symmetry operation of the space group of the parent lattice;

\item  all force constants $k_{\sigma _{i}\sigma _{j}}$ are such that 
\begin{equation}
\frac{k_{\sigma _{i}\sigma _{j}}}{\sqrt{k_{\sigma _{i}\sigma _{i}}k_{\sigma
_{j}\sigma _{j}}}}\ll 1. 
\end{equation}
\end{itemize}

Consider a $d$-dimensional solid made of $N$ atoms connected by springs of
characterized by symmetrically equivalent tensors $k\phi \left( i,j\right) $%
. Without loss of generality, the masses of all atoms are set to unity since
the formation entropies in the high temperature limit are independent of the
atomic masses (see Appendix \ref{masscancel}). In the high temperature
limit, the vibrational free energy per atom is given by: 
\begin{equation}
F_{\text{vib}}=\frac{k_{B}T}{2N}\sum_{m}\ln \lambda _{m}  \label{svib}
\end{equation}
where the sum is taken over the nonzero eigenvalues $\lambda _{m}$ of the
dynamical matrix $D$ of the system. (The zero eigenvalues correspond the
modes where the whole crystal moves rigidly. In the thermodynamic limit,
these few missing degrees of freedom are inconsequential.)

Because all springs in the system are equivalent to each other, matrix $D$
can be written as 
\begin{equation}
D=kC 
\end{equation}
where $C$ is a matrix of dimensionless geometrical factors independent of $k$
but specific to the type of lattice. From this expression of $D$, it follows
naturally that eigenvectors of $D$ are independent of $k$ and that its
eigenvalues can be written as 
\begin{equation}
\lambda _{m}=kl_{m} 
\end{equation}
where the $l_{m}$ are geometric factors independent of $k$.

Consider what happens to $S_{\text{vib}}$ when the stiffness of one of the
springs is changed from $k$ to $k+\Delta k$. Let $\Delta D$ denote the
corresponding change in matrix $D$. To the first order, the resulting
changes in the eigenvalues are given by: 
\begin{equation}
\Delta \lambda _{m}=u_{m}^{\prime }\Delta Du_{m} 
\end{equation}
where $u_{m}$ is the (dimensionless) eigenvector of $D$ associated with
eigenvalue $\lambda _{m}$. Since $D$ is linear in the spring constants, we
can write 
\begin{equation}
\Delta D=\Delta k\;B 
\end{equation}
where $B$ is matrix of geometrical factors independent of $k$ and $\Delta k$
but specific to the type of lattice. While matrix $B$ also depends on which
spring is being modified, the matrices $B$ corresponding to each spring are
equivalent under a symmetry operation of the crystal's space group. The
changes in the eigenvalues can then be expressed as: 
\begin{eqnarray*}
\Delta \lambda _{i} &=&\Delta ku_{m}^{\prime }Bu_{m} \\
&\equiv &\Delta k\;g_{m}
\end{eqnarray*}
where $g_{i}$ is a dimensionless number independent of $k$ and $\Delta k$.

Substituting these results into Equation (\ref{svib}), we obtain: 
\begin{equation}
F_{\text{vib}}=\frac{k_{B}T}{2N}\sum_{m}\ln \left( kl_{m}+\Delta
k\;g_{m}\right) . 
\end{equation}
To the first order, we can express the vibrational entropy change as 
\begin{eqnarray*}
\Delta F_{\text{vib}} &=&{\left. \frac{\partial F_{\text{vib}}}{\partial
\Delta k}\right| }_{\Delta k=0}\Delta k \\
&=&\frac{k_{B}T}{2N}\sum_{m}\frac{g_{m}}{kl_{m}}\Delta k \\
&=&k_{B}T\left( \frac{1}{2N}\sum_{m}\frac{g_{m}}{l_{m}}\right) \frac{\Delta k%
}{k} \\
&\equiv &k_{B}TG\frac{\Delta k}{k}.
\end{eqnarray*}
where $G$ is a dimensionless geometrical factor depending only on the
lattice type.

In the limit of $\Delta k\ll k$, we can obtain the change in vibrational
entropy due to a change in all the spring constants by simply summing the
effect of the change $\Delta k_{s}$ in the stiffness of each spring $s$: 
\begin{equation}
\Delta F_{\text{vib}}=k_{B}TG\sum_{s}\frac{\Delta k_{s}}{k}  \label{dS1}
\end{equation}
To determine the value of $G$, we consider the following particular case for
which the exact vibrational entropy change is known. Once the value of $G$
is known, it can be used in any other case sharing a same lattice type.

In a solid bound by springs of stiffness $k$ is given by, if the stiffness
of all springs is increased by $\Delta k$, each eigenvalue $\lambda _{m}$
becomes $\lambda _{m}\frac{k+\Delta k}{k}$ and the vibrational entropy
becomes: 
\begin{eqnarray*}
F_{\text{vib}}^{\prime } &=&\frac{k_{B}T}{2N}\sum_{i}\ln \left( \lambda _{i}%
\frac{k+\Delta k}{k}\right) \\
&=&\frac{k_{B}T}{2N}\sum_{i}\left( \ln \lambda _{i}+\ln \frac{k+\Delta k}{k}%
\right) \\
&=&F_{\text{vib}}+\frac{k_{B}T}{2N}\sum_{i}\ln \frac{k+\Delta k}{k} \\
&\approx &F_{\text{vib}}+k_{B}T\frac{Nd}{2N}\frac{\Delta k}{k}+O\left(
\left( \Delta k\right) ^{2}\right) \\
&=&F_{\text{vib}}+k_{B}T\frac{Nd}{2N}\frac{1}{ZN/2}\sum_{s}\frac{\Delta k}{k}
\\
&=&F_{\text{vib}}+\frac{k_{B}Td}{ZN}\sum_{s}\frac{\Delta k}{k}.
\end{eqnarray*}
where $Z$ is the number of nearest neighbors and $\sum_{s}$ denotes a sum
over all nearest neighbor bonds. Since this result is exact to the first
order, we can compare it to Equation (\ref{dS1}) and identify the unknown
constant $G$ to be $\frac{d}{ZN}$. We thus obtain the following result: 
\begin{equation}
\Delta F_{\text{vib}}=\frac{3k_{B}T}{ZN}\sum_{s}\frac{\Delta k_{s}}{k}.
\label{dvibo1}
\end{equation}

We now turn to the problem of calculating the vibrational entropy of mixing
in a binary alloy. We first define a normalized $3N\times 3N$ dynamical
matrix $\hat{D}$ as follows: 
\begin{equation}
\hat{D}_{\alpha \beta }\left( i,j\right) =\frac{\Phi _{\alpha \beta }\left(
i,j\right) }{\sqrt{k_{\sigma _{i}\sigma _{i}}k_{\sigma _{j}\sigma _{j}}}} 
\end{equation}
where $k_{\sigma _{i}\sigma _{i}}$ is the spring constant of an $A-A$ bond
if site $i$ is occupied by a $A$ atom similarly for a site occupied by a $B$
atom. For the purpose of calculating free energy of formation, this
normalized dynamical matrix gives the same result as the usual dynamical
matrix because the factors in the denominator exactly cancel out, for the
same reason masses cancel out (See Appendix \ref{masscancel}). This
transformation normalizes the spring constant associated with $A-A$ bonds
and $B-B$ bonds to $1$ while the spring constant associated with $A-B$ bond
becomes $\left( k_{AB_{j}}/\sqrt{k_{AA}k_{BB}}\right) $ where $k_{AB}$, $%
k_{AA}$ and $k_{BB}$ respectively denote the true spring constants of $A-B$, 
$A-A$ and $B-B$ bonds. The usefulness of this normalization is to extend the
applicability of Equation (\ref{dvibo1}) to the case where $k_{AA}$ and $%
k_{BB}$ are very different.

Let us start with a phase separated mixture of $A$ and $B$ atoms. Let us
think of this system as one where all atoms are identical but where the
springs connecting them can be either one of three types $A-A$, $B-B$ or $%
A-B $. Where the springs are placed defines which type of atom sits at each
site. We now replace one $A-A$ bond in the pure $A$ phase by an $A-B$ bond
and one $B-B$ bond in the pure $B$ phase by an $A-B$ bond. By Equation (\ref
{dvibo1}), the resulting change in vibrational entropy per atom is: 
\begin{equation}
\Delta F_{\text{vib}}=\frac{k_{B}Td}{ZN}\left( \frac{k_{AB}}{\sqrt{%
k_{AA}k_{BB}}}-1+\frac{k_{AB}}{\sqrt{k_{AA}k_{BB}}}-1\right) . 
\end{equation}
To satisfy the assumptions of the above derivation, we require that $k_{AB}/%
\sqrt{k_{AA}k_{BB}}\ll 1$. If we create a total number $n_{AB}$ of $A-B$
bonds, we perform the above operation $n_{AB}/2$ times and the vibrational
entropy change is: 
\begin{equation}
\Delta F_{\text{vib}}=\frac{n_{AB}}{2}\frac{k_{B}Td}{ZN}\left( \frac{k_{AB}}{%
\sqrt{k_{AA}k_{BB}}}-1+\frac{k_{AB}}{\sqrt{k_{AA}k_{BB}}}-1\right) 
\end{equation}
To the first order (when $k_{AB}/\sqrt{k_{AA}k_{BB}}\ll 1$), this expression
is equivalent to 
\begin{equation}
\Delta F_{\text{vib}}=\frac{n_{AB}}{N}\frac{k_{B}Td}{2Z}\ln \left( \frac{%
k_{AB}^{2}}{k_{AA}k_{BB}}\right) . 
\end{equation}
The nearest neighbor ECI of the cluster expansion of the vibrational free
energy is thus: 
\begin{equation}
V_{1nn}=\frac{d}{8}k_{B}T\ln \left( \frac{k_{AA}k_{BB}}{k_{AB}^{2}}\right) . 
\end{equation}

\section{Instability}

\label{instab}An extreme case of anharmonicity occurs when the energy
surface, in the neighborhood of a configuration ${\bf\sigma}$, has no local
minimum. As noted in \cite{craievich:siep} and \cite{craievich:nicr}, this
situation occurs sufficiently frequently to deserve a particular attention.
A typical example of such a situation occurs when the fcc-based $L1_{0}$
structure is unstable with respect to a deformation along the Bain path,
which leads to a bcc-based $B2$ structure. While it is possible to construct
a separate cluster expansion for the fcc and bcc phases, the fundamental
question that arises is: What is the free energy of the $L1_{0}$ structure?
Since it is unstable, the standard harmonic expression for the free energy
can clearly not be used.

One suggested solution to this problem, described in \cite{craievich:nicr},
is to perform the coarse graining in a different order than presented in
Section \ref{coarse}. The sum over configurations is performed first, and
the vibrational properties of the configurational averaged alloy are then
calculated. The main limitation of this approach is that it would be
extremely difficult to compute the averaged vibrational properties by any
other method than by the so-called virtual crystal approximation (see
Section \ref{appdisord}). Another limitation is that it only addresses
instabilities with respect to cell shape distortions, ignoring instabilities
with respect to internal degrees of freedom ({\it i.e.} atomic positions).

In this section, we present another approach to solve the instability
problem. We argue that the general formalism developed in Section \ref
{coarse} can in fact be adapted to allow for instability.

While the coarse graining technique is most naturally interpreted as
integrating out the ``fast'' degrees of freedom ({\it e.g.} vibrations)
before considering ``slower'' ones ({\it e.g.} configurational changes)\cite
{ceder:ising}, the time scale of the various types of excitations is, in
fact, irrelevant. The partition function is simply a sum over states which
can be calculated in any order. As long as we can associate any vibrational
state $v$ of the system with a configuration ${\bf\sigma}$, the coarse
graining procedure remains valid.

Under this point of view, it is clear that it does not matter whether there
is even a local minimum of energy in the portion of phase space associated
with configuration ${\bf\sigma}$. What is important, however, is that the
neighborhood of configuration ${\bf\sigma}$ in phase space is thoroughly
sampled ({\it i.e.} that the constrained system is ergodic) over a
macroscopic time scale. There is no need for ergodicity within the time
scale of the configurational excitations. If the neighborhood of a given
configuration ${\bf\sigma}$ is not fully sampled before the alloy jumps to
another configuration ${\bf\sigma}^{\prime }$, it is still possible that
the unsampled portion of phase space around ${\bf\sigma}$ will be visited
at a later time, when the system returns to the neighborhood of
configuration ${\bf\sigma}$. The ergodicity requirement at the macroscopic
time scale imposes the important but intuitively obvious constraint that the
phase space neighborhood of configuration ${\bf\sigma}$ cannot contain
states that are associated to different phases of the system.

This discussion shows that there is no fundamental limitation to the
applicability of the standard coarse graining framework in the presence of
instability. However, we still need to describe how the free energy of an
unstable configuration could be determined in practice. The task is
simplified by the fact that the free energy of an unstable stable does not
need to be extremely accurately determined, because unstable states are
relatively rarely visited, even at high temperatures. Nevertheless, it is
important to assign a free energy to those unstable states, to ensure that
the Ising model used to represent the alloy is well-defined.

The free energy associated with one configuration can be obtained by
integrating $\exp \left[ {-\beta E({\bf\sigma},v)}\right] $ with respect to 
$v$ over the portion of phase space associated with ${\bf\sigma}$. In the
classical limit, we can label the vibrational states $v$ by the position
each particle takes and the integration limits can be found by geometrical
arguments. The quantum mechanical equivalent of this operation is complex,%
\footnote{%
The quantum partition function can be written as the trace of the matrix $%
\exp \left( -\beta H\right) $, where $H$ is the (multibody) Hamiltonian of
the system. The trace can computed in any convenient basis and in particular
one could use Dirac delta functions. In this fashion, it is possible to
define a localized free energy by summing only over the delta functions
located in the neighborhood of one configuration ${\bf\sigma}$.} but
unlikely to be needed in practice. The unstable states are essentially never
visited at low temperatures, where a quantum mechanical treatment would be
essential \footnote{%
This observation is related to the fact that quasi-harmonic approximation,
which allows a quantum-mechanical treatment, is accurate up to a temperature
where the classical limit is reached.}.

Focusing on the classical limit, we consider an unstable configuration ${\bf%
\sigma}$. Let $D$ be the dynamical matrix evaluated at the saddle point of
the energy surface closest to the ideal undistorted configuration ${\bf%
\sigma}$.\footnote{%
Is is possible that an unstable configuration ${\bf\sigma}$ cannot be
associated with a saddle point and the derivation would have to be modified.
In particular the bounds of integration would have to be made asymmetric.}
We consider that when the state $v$ of the system is such that one atom
moves away from its position at the saddle point by more than $r$, it should
not longer be considered part of configuration ${\bf\sigma}$. For an
instability with respect to internal degrees of freedom (atomic positions),
a natural choice for $r$ would be half the average nearest neighbor
interatomic distance. For an instability with respect to unit cell
deformation, $r$ could be half the change in the average nearest neighbor
distance induced by the displacive transformation.

The boundedness of the portion of phase space associated with ${\bf\sigma}$
can be accounted for by replacing the usual classical partition function
associated with one normal mode of oscillation $i$ by 
\begin{equation}
\frac{1}{h}\int_{-L_{i}}^{L_{i}}\exp \left( -\frac{1}{2}\beta \,\dot{s}%
^{2}\right) d\dot{s}\int_{-L_{i}}^{L_{i}}\exp \left( -\frac{1}{2}\beta
\lambda _{i}s^{2}\right) ds 
\end{equation}
where $\lambda _{i}$ is the $i$-th eigenvalue of the dynamical matrix, $h$
is Planck's constant and $L_{i}$ is a measure of the size of the phase space
neighborhood of ${\bf\sigma}$ along the direction associated with normal
mode $i$. This size parameter can be expressed in terms of the parameter $r$
just introduced. Let $u_{i}\left( j\right) =\frac{e_{i}\left( j\right) }{%
\sqrt{M_{j}}}$ where $e_{i}$ is the $i$-th eigenvector of $D$ and $M_{j}$ is
the mass of atom $j$. After normalizing $u_{i}$ so that $\sum_{j}u_{i}^{2}%
\left( j\right) =N$, the number of atom in the system, we can then write 
\begin{equation}
L_{i}=r\left( \max_{\text{nn }j,j^{\prime }}\left\| u_{i}\left( j\right)
-u_{i}\left( j^{\prime }\right) \right\| \right) ^{-1} 
\end{equation}
where the maximum is taken over all nearest neighbor pairs of atoms $%
j,j^{\prime }$. This choice of integration bounds approximately defines a
neighborhood of ${\bf\sigma}$ such that no atom moves farther than $r$ from
its position at the saddle point (relative to its neighbors). In this
approximation, the free energy of an unstable state is given by: 
\begin{equation}
\frac{F}{N}=\frac{E^{\ast }}{N}-\frac{k_{B}T}{N}\sum_{i}\ln \left( \frac{%
k_{B}T}{h\nu _{i}}\mbox{erf}\left( L_{i}\sqrt{\frac{2\left( \pi \nu
_{i}\right) ^{2}}{k_{B}T}}\right) \right) 
\end{equation}
where $\nu _{i}$ is the frequency of normal mode $i$ and where the error
function for real or imaginary arguments is given by 
\begin{equation}
\mbox{erf}\left( u\right) =\frac{2u}{\sqrt{\pi }}%
\int_{0}^{1}e^{-u^{2}s^{2}}ds. 
\end{equation}
The suggested definition of the free energy of an unstable configuration has
interesting properties. First, as the neighborhood size $L_{i}$ increases,
the expression reduces to the usual harmonic expression. The effect of the
correction is not limited to unstable modes: Modes that are so soft that it
is likely that the motion of the atoms exceeds $r$ are also affected. There
may obviously be other definitions of $L_{i}$.\ The above example simply
gives an example of how it could be calculated.

Going back to our initial example of the $L1_{0}\rightarrow B2$ instability,
we can now outline how this problem could be handled within the traditional
coarse graining scheme. Two separate clusters expansion need to be
constructed, one for the bcc phases and one for the fcc phases. But since we
now know how to assign a free energy to the unstable $L1_{0}$ configuration,
the fcc cluster expansion can be successfully defined. The free energy
attributed to the $L1_{0}$ configuration should be sufficiently high so that
the free energy curve of the fcc phase in the vicinity of 0.5 concentration
will lie above the free energy curve of the bcc phase, as it should. The
fact that both CVM or Monte Carlo calculations on the fcc lattice would
attribute a positive probability to $L1_{0}$-like structures should not be
regarded as a problem: This is precisely what will ensure that the
calculated fcc free energy curve lies above the bcc one.

The discussion has so far been concerned with the expression of the
partition function, which is the relevant quantity to consider when the
phase diagram is calculated with the CVM or the low temperature expansion.
Let us now consider the implications of this approach to Monte-Carlo
simulations. Thermodynamic quantities derived from averages, such as the
average energy, are obviously unaffected by the presence of unstable
configurations. For quantities derived from fluctuations, such as the heat
capacity, slight modifications are needed. In traditional Monte Carlo
simulations, the heat capacity arising from vibrational degrees of freedom
is consistently neglected, and any thermodynamic quantity obtained from
Monte Carlo can be unambiguously interpreted as the configurational
contribution. In the more general setting presented here, there is an
overlap between vibrational and configurational fluctuations and the only
way to obtain well defined thermodynamic quantities is to fully account for
the vibrational fluctuations. Fortunately, there is a straightforward way to
do so. The total variance of the energy (or any other quantity) can be
exactly expressed as a sum of the variance within each configuration ${\bf%
\sigma}$ and the variance of the average energy of each configuration: 
\begin{multline*}
\left\langle E^{2}\right\rangle -\left\langle E\right\rangle ^{2}=\sum_{{\bf%
\sigma}}\sum_{v\in {\bf\sigma}}P_{{\bf\sigma}v}E_{{\bf\sigma}%
v}^{2}-\left( \sum_{{\bf\sigma}}\sum_{v\in {\bf\sigma}}P_{{\bf\sigma}v}E_{%
{\bf\sigma}v}\right) ^{2} \\
=\left( \sum_{{\bf\sigma}}P_{{\bf\sigma}}\overline{E}_{{\bf\sigma}%
}^{2}-\left( \sum_{{\bf\sigma}}P_{{\bf\sigma}}\overline{E}_{{\bf\sigma}%
}\right) ^{2}\right) +\sum_{{\bf\sigma}}P_{{\bf\sigma}}\left( \sum_{v\in 
{\bf\sigma}}\frac{P_{{\bf\sigma}v}}{P_{{\bf\sigma}}}\left( E_{{\bf\sigma}%
v}-\overline{E}_{{\bf\sigma}}\right) ^{2}\right) .
\end{multline*}
where $P_{{\bf\sigma}v}$ is the probability of finding the system in state $%
{\bf\sigma}$,$v$ while $P_{{\bf\sigma}}=\sum_{v\in {\bf\sigma}}P_{{\bf%
\sigma}v}$ and $\overline{E}_{{\bf\sigma}}=\sum_{v\in {\bf\sigma}}\frac{P_{%
{\bf\sigma}v}}{P_{{\bf\sigma}}}E_{{\bf\sigma}v}$. The first term is the
usual fluctuation obtained from Monte Carlo. The second term is a correction
which takes the form of a simple configuration average of fluctuations
within each configuration. The fluctuation of a system constrained to remain
in the vicinity of configuration ${\bf\sigma}$ is usually just as simple to
determine as its average properties. In the case of energy, the fluctuations
within each configuration are simply related the heat capacity of a harmonic
solid.

The main objective of this section was to show that there is no fundamental
problem associated with unstable states in coarse graining formalism. While
it is true that the free energy of an unstable configuration is not uniquely
defined, once a particular way to coarse grain phase space is chosen, the
free energy of all configurations can be defined in a consistent fashion.
There are admittedly some practical issues to be resolved regarding the
practical implementation of coarse graining in the presence of
instabilities, but the approach suggested in this section indicates that
these difficulties can be overcome.



\begin{table}
\begin{tabular}{llrrllc}
Composition & Transition & \multicolumn{1}{c}{$\Delta S$} & 
\multicolumn{1}{c}{$T$} & Methods & Refs. \\ 
&  & \multicolumn{1}{c}{$\left( \frac{k_{B}}{\mbox{atom}} \right)$} & 
\multicolumn{1}{c}{(K)} &  &  \\ \hline
AgCu & L1$_{0}^{\ast }$ (form) & -0.11 & high & ASW,MJS & \cite{sanchez:agcu}
$^{\dagger }$ \\ 
Ag$_{3}$Cu & L1$_{2}^{\ast }$ (form) & -0.22 & high & ASW,MJS & \cite
{sanchez:agcu} $^{\dagger }$ \\ 
Ag$_{3}$Cu & L1$_{2}^{\ast }$ (form) & -0.02 & high & ASW,MJS & \cite
{sanchez:agcu} $^{\dagger }$ \\ 
Cu$_{3}$Au & L1$_{2}$ $\rightarrow $ fcc rnd & 0.12 & 663 & TB,BvK,QH,VCA & 
\cite{cleri:cu3au} \\ 
CdMg & hcp rnd (form) & 0.13 & 900 & LMTO,MJS,CE & \cite{asta:cdmg} \\ 
CdMg & B19 (form) & 0.14 & high & LMTO,MJS & \cite{asta:cdmg} $^{\dagger } $ \\ 
Cd$_{3}$Mg & D0$_{19}$ (form) & -0.03 & high & LMTO,MJS & \cite{asta:cdmg} $%
^{\dagger }$ \\ 
CdMg$_{3}$ & D0$_{19}$ (form) & -0.10 & high & LMTO,MJS & \cite{asta:cdmg} $%
^{\dagger }$ \\ 
Cu$_{3}$Au & L1$_{2}$ $\rightarrow $ fcc rnd$^{\ast }$ & 0.10 & high & 
EAM,BvK,H,SC & \cite{ackland:vol} \\ 
Ni$_{3}$Al & L1$_{2}$ $\rightarrow $ fcc rnd$^{\ast }$ & 0.29 & high & 
EAM,BvK,H,SC & \cite{ackland:vol} \\ 
SiGe & B3$^{\ast }$ (form) & -0.02 & high & PP,BvK,H & \cite{gdg:phd} \\ 
ArKr & L1$_{0}$ (form) & -0.06 & high & pot.,BvK,H & \cite{gdg:phd} \\ 
Ca$_{0.5}$Mg$_{0.5}$O & fcc rnd (form) & 0.04 & high & SCPIB,BvK,H,CE & \cite
{tepesch:caomgo} $^{\dagger }$ \\ 
Ni$_{3}$Al & L1$_{2}$ $\rightarrow $ fcc rnd$^{\ast }$ & 0.27 & 1400 & 
EAM,BvK,QH,SC & \cite{althoff:vib} \\ 
NiCr & fcc rnd (form) & n.a. & 1550 & FLASTO,VCA & \cite{craievich:nicr}  \\ 
Ni$_{3}$Al & L1$_{2}$ $\rightarrow $ fcc rnd$^{\ast }$ & 0.20 & 1500 & 
EAM,MC,SC & \cite{morgan:phd} \\ 
Ni$_{3}$Al & L1$_{2}$ $\rightarrow $ fcc rnd$^{\ast }$ & 0.22 & 1200 & 
EAM,MD,SC & \cite{ravelo:vib} \\ 
Ni$_{3}$Al & L1$_{2}$ $\rightarrow $ fcc rnd$^{\ast }$ & 0.00 & high & 
PP,BvK,QH,SQS & \cite{avdw:ni3al} \\ 
Ni$_{3}$Al & L1$_{2}$ $\rightarrow $ D0$_{22}$ & 0.04 & high & PP,BvK,QH & 
\cite{avdw:ni3al} \\ 
CuAu & L1$_{0}$ $\rightarrow $ fcc rnd & 0.18 & 800 & PP,LR,QH,CE & \cite
{ozolins:cuau} \\ 
Cu$_{3}$Au & L1$_{2}$ $\rightarrow $ fcc rnd & 0.08 & 800 & PP,LR,QH,CE & 
\cite{ozolins:cuau} \\ 
CuAu$_{3}$ & L1$_{2}$ $\rightarrow $ fcc rnd & 0.05 & 800 & PP,LR,QH,CE & 
\cite{ozolins:cuau} \\ 
CuAu & L1$_{1}^{\ast }$ (form) & 0.58 & 800 & PP,LR,QH & \cite{ozolins:cuau}
\\ 
CuAu & L1$_{0}$ (form) & 0.21 & 800 & PP,LR,QH & \cite{ozolins:cuau} \\ 
Cu$_{3}$Au & L1$_{2}$ (form) & 0.20 & 800 & PP,LR,QH & \cite{ozolins:cuau}  \\ 
CuAu$_{3}$ & L1$_{2}$ (form) & 0.26 & 800 & PP,LR,QH & \cite{ozolins:cuau}  \\ 
Al$_{3}$Li & L1$_{2}^{\ast }$ $\rightarrow $ D0$_{22}^{\ast }$ & 0.04 & high
& PP,BvK,H & \cite{sluiter:cafc} \\ 
Fe$_{3}$Al & D0$_{3}$ $\rightarrow $ bcc rnd & 0.11 & high & 
TB-LMTO,pot.,H,SC & \cite{shaojun:feal} \\ 
Pd$_{3}$V & D0$_{22}$ $\rightarrow $ fcc rnd & -0.07 & high & PP,BvK,QH,SQS
& \cite{avdw:pd3v} \\ 
Pd$_{3}$V & L1$_{2}$ $\rightarrow $ D0$_{22}$ & 0.08 & high & PP,BvK,QH & 
\cite{avdw:pd3v} \\
Al$_3$Sc & (form) & -0.70 & high & PP,LR,QH & \cite{ozolins:alsc}\\
Al$_{26}$Sc & (form, per Sc atom) & 0.50 & high & PP,LR,QH & \cite{ozolins:alsc}\\
Al$_2$Cu & $\theta'_c \rightarrow theta$ & 0.37 & high & PP,LR,H & \cite{wolverton:cual}
\end{tabular}

(form): Vibrational entropy of formation from pure elements.

rnd: Disordered solid solution.

$\ast $: metastable compound

$\dagger $: calculated from the data presented in the paper.

\caption[Calculated Vibrational Entropy Differences.]{Calculated Vibrational Entropy Differences. See Table \ref{methabb}
for abbreviations.\label{tabsvibth}}%
\end{table}

\begin{table}
\begin{center}
\begin{tabular}{ll}
pot.: pair potentials rmonic approximation \\ 
EAM: embedded atom method & QH: quasiharmonic approximation \\ 
TB: tight-binding & MC: Monte Carlo \\ 
TB-LMTO: tight-binding LMTO & MD: molecular dynamics \\ 
LMTO: linear muffin-tin orbitals & D: Debye model \\ 
ASW: atomic spherical waves & BvK: Born-von K\'{a}rm\'{a}n spring model \\ 
PP: pseudopotential calculations & SC: supercell method \\ 
& LR: linear response \\ 
cal.: differential calorimetry measurements & VCA: virtual crystal
approximation \\ 
1xtal: single crystal phonon dispersion measurements & MJS: Moruzzi, Janak
\& Schwarz model \\ 
INS: incoherent neutron scattering measurements & CE: cluster expansion \\ 
anh.: anharmonicity included. & SQS: special quasirandom structures
\end{tabular}
\end{center}

\caption{Abbreviations Used in Tables \ref{tabsvibth} and
\ref{tabsvibexp}\label{methabb}.}%
\end{table}

\begin{table}[tbp]

\begin{center}
\begin{tabular}{llrrllc}
Composition & Transition & \multicolumn{1}{c}{$\Delta S$} & 
\multicolumn{1}{c}{$T$} & Methods & Refs. \\ 
&  & \multicolumn{1}{c}{$\left(\frac{k_{B}}{\mbox{atom}}\right)$} & 
\multicolumn{1}{c}{(K)} &  &  \\ \hline
Ni$_{3}$Al & L1$_{2}$ $\rightarrow $ fcc rnd & $0.27$\phantom{$\mbox{}%
\pm0.00$} & high & cal.,D,H & \cite{fultz:ni3alcal} \\ 
Ni$_{3}$Al & L1$_{2}$ $\rightarrow $ fcc rnd & $0.19$\phantom{$\mbox{}%
\pm0.00$} & 343 & cal. & \cite{fultz:ni3alcal} \\ 
Fe$_{3}$Al & D0$_{3}$ $\rightarrow $ bcc rnd & $0.10\pm0.03$ & high & 
cal.,D,H & \cite{fultz:fe3al2} \\ 
Cu$_{3}$Al & L1$_{2}$ $\rightarrow $ fcc rnd & $0.14\pm0.05$ & high & 
cal.,1xtal,H,VCA & \cite{fultz:cu3au} \\ 
Fe$_{0.70}$Cr$_{0.30}$ & bcc rnd (form) & $0.14\pm0.05$ & high & 1xtal,H,VCA
& \cite{fultz:fecr} \\ 
Fe$_{0.53}$Cr$_{0.47}$ & bcc rnd (form) & $0.20\pm0.05$ & high & 1xtal,H,VCA
& \cite{fultz:fecr} \\ 
Fe$_{0.30}$Cr$_{0.70}$ & bcc rnd (form) & $0.21\pm0.05$ & high & 1xtal,H,VCA
& \cite{fultz:fecr} \\ 
Ni$_{3}$Al & L1$_{2}^{\ast }$ $\rightarrow $ fcc rnd & $0.10$%
\phantom{$\mbox{}\pm0.00$} & high & INS,H & \cite{fultz:ni3alphon} \\ 
Ni$_{3}$Al & L1$_{2}^{\ast }$ $\rightarrow $ fcc rnd & $0.30$%
\phantom{$\mbox{}\pm0.00$} & high & INS,H,VCA & \cite{fultz:ni3alphon} \\ 
Ni$_{3}$V & D0$_{22}$ $\rightarrow $ fcc rnd & $0.04\pm0.02$ & 300 & cal.,INS
& \cite{fultz:ni3v} \\ 
Co$_{3}$V & L1$_{2}^{\ast }$ $\rightarrow $ fcc rnd & $0.15\pm0.02$ & high & 
INS & \cite{fultz:co3v} \\ 
Cu$_{3}$Au & L1$_{2}$ (form) & $0.06\pm0.03$ & 300 & INS,anh. & \cite
{fultz:cu3au2} \\ 
Cu$_{3}$Au & L1$_{2}$ (form) & $0.12\pm0.03$ & 800 & INS,anh. & \cite
{fultz:cu3au2} \\ 
Co$_{3}$V & hP24 $\rightarrow $ fcc rnd & $0.07$\phantom{$\mbox{}\pm0.00$} & 
high & INS,QH & \cite{fultz:co3v2} \\ 
CeSn$_{3}$ & $\gamma $-Ce + $\beta $-Sn $\rightarrow $ L1$_{2}$ & $%
-0.54\pm0.09$ & high & 1xtal,H & \cite{fultz:trends} \\ 
LaSn$_{3}$ & hcp-La + $\beta $-Sn $\rightarrow $ L1$_{2}$ & $-0.43\pm0.09$ & 
high & 1xtal,H & \cite{fultz:trends} \\ 
Ni$_{3}$Al & L1$_{2}$ (form) & $-0.20\pm0.03$ & high & 1xtal,H & \cite
{fultz:trends} \\ 
Ni$_{3}$Fe & fcc-Ni + bcc-Fe $\rightarrow $ L1$_{2}$ & $0.09\pm0.03$ & high
& 1xtal,H & \cite{fultz:trends} \\ 
Pt$_{3}$Fe & fcc-Pt + bcc-Fe $\rightarrow $ L1$_{2}$ & $0.14\pm0.03$ & high
& 1xtal,H & \cite{fultz:trends} \\ 
Pd$_{3}$Fe & fcc-Pd + bcc-Fe $\rightarrow $ L1$_{2}$ & $0.05\pm0.03$ & high
& 1xtal,H & \cite{fultz:trends} \\ 
Cu$_{3}$Zn & fcc-Cu + hcp-Zn $\rightarrow $ L1$_{2}$ & $-0.01\pm0.03$ & high
& 1xtal,H & \cite{fultz:trends} \\ 
Cu$_{3}$Au & L1$_{2}$ (form) & $0.07\pm0.03$ & high & 1xtal,H & \cite
{fultz:trends} \\ 
Fe$_{3}$Pt & bcc-Fe + fcc-Pt $\rightarrow $ L1$_{2}$ & $0.55\pm0.03$ & high
& 1xtal,H & \cite{fultz:trends} \\ 
Fe$_{3}$Al & bcc-Fe + fcc-Al $\rightarrow $ D0$_{3}$ & $-0.06\pm0.03$ & high
& 1xtal,H & \cite{fultz:trends}
\end{tabular}
\end{center}

$\ast $: metastable compound

\caption[Experimental Measurements of Vibrational Entropy Differences.]{Experimental
Measurements of Vibrational Entropy Differences. See
Table \ref{methabb} for abbreviations.\label{tabsvibexp}}%
\end{table}

\begin{table}
\begin{center}
\begin{tabular}{llll}
$T$ ($K$) & $S_{vib}^{o\rightarrow d}$ & $\Delta V^{o\rightarrow d}/V^{o}$
(\%) ence \\ \hline
high & 0.29 & 3\% & \cite{ackland:vol} \\ 
1200 & 0.22 & 2\% & \cite{ravelo:vib} \\ 
1000 & 0.15 & 2\% & \cite{althoff:vib}\cite{althoff:vibcms} \\ 
1000 & 0.11 & 1.6\% & \cite{morgan:phd}, Foiles-Daw EAM potentials \\ 
high & 0.00 & 0.5\% & \cite{avdw:ni3al}
\end{tabular}
\end{center}

\caption{Relation Between the Vibrational Entropy Change upon Disordering
and the Volume Change upon Disordering in Various Theoretical
Investigations of the Ni$_{3}$Al Compounds \label{ni3alSV}} 
\end{table}

\begin{table}[tbp]
\begin{center}
\begin{tabular}{ccc}
Compound (Structure) & ``Stiffness vs. Length'' Model & 1nn Spring Model \\ 
\hline
Pd$_3$V L1$_{2}$ & -4.39 & -4.39 \\ 
Pd$_3$V D0$_{22}$ & -4.42 & -4.47 \\ 
Pd$_3$V SQS-8 & -4.56 & -4.54 \\ 
Ni$_3$Al L1$_{2}$ & -5.57 & -5.55 \\ 
Ni$_3$Al SQS-8 & -5.54 & -5.57
\end{tabular}
\end{center}
\caption{Comparison Between Vibrational Entropies Obtained from the
``Bond Stiffness vs. Bond Length'' Model and from a First Nearest Neighbor Spring
Model.}
\label{svslpred}
\end{table}

\illuseps{3.4in}{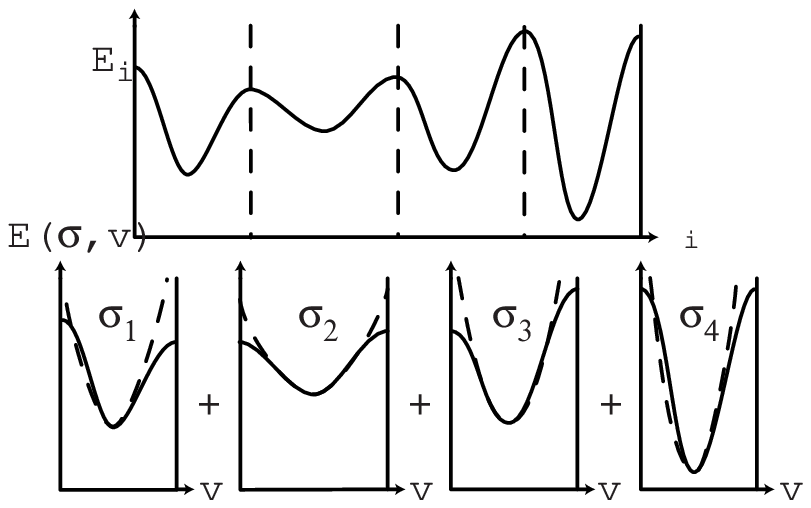}{The Coarse Graining Approach.
The global energy surface of an alloy system, which gives the 
energy $E_i$ of each state $i$, can be partitioned into
a set of local energy surfaces, each of which is associated with 
a distinct configuration $\sigma_k$ of an Ising model.
Within each configuration $\sigma_k$, the local energy surface
gives the energy as a function of the atomic displacements $v$
(or any other nonconfigurational degrees of freedom).
The thermodynamics of the system thus reduces
to the one of an traditional Ising model, where the ``energy''
of each configuration now becomes the free energy associated
with the corresponding local energy surface.
If the local energy surfaces can be considered quadratic,
their associated free energy can be determined by
solving a Born-von K\'{a}rm\'{a}n phonon problem.
}

\illuseps{3.4in}{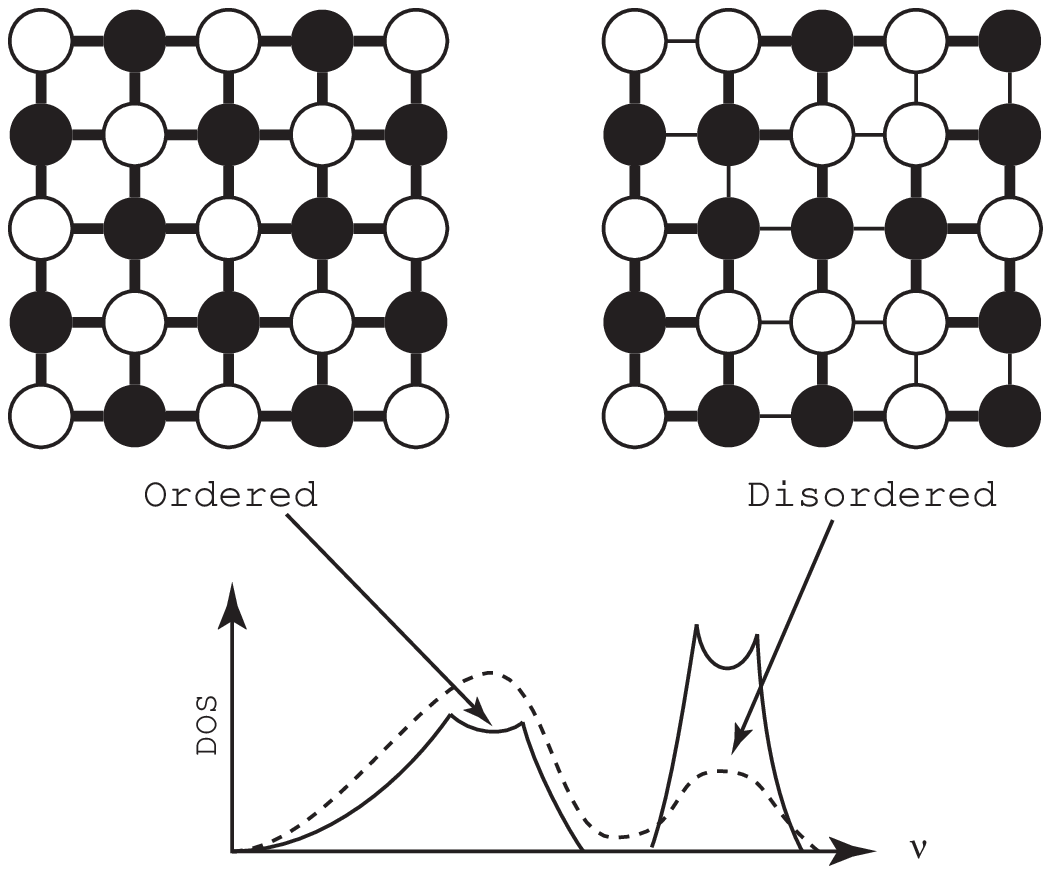}{The ``Bond Proportion'' Mechanism.
Ordered alloys are characterized by the fact that
a large proportion of the nearest-neighbor bonds
join unlike atoms.
As these types of bonds are presumably stiffer, they are
responsible for an increased density of high-frequency
optical phonon modes. Disordering reduces the proportion of stiff
nearest-neighbor bonds, and the density of high frequency optical
modes decreases correspondingly, resulting in an increase in
vibrational entropy.
}

\illuseps{3.4in}{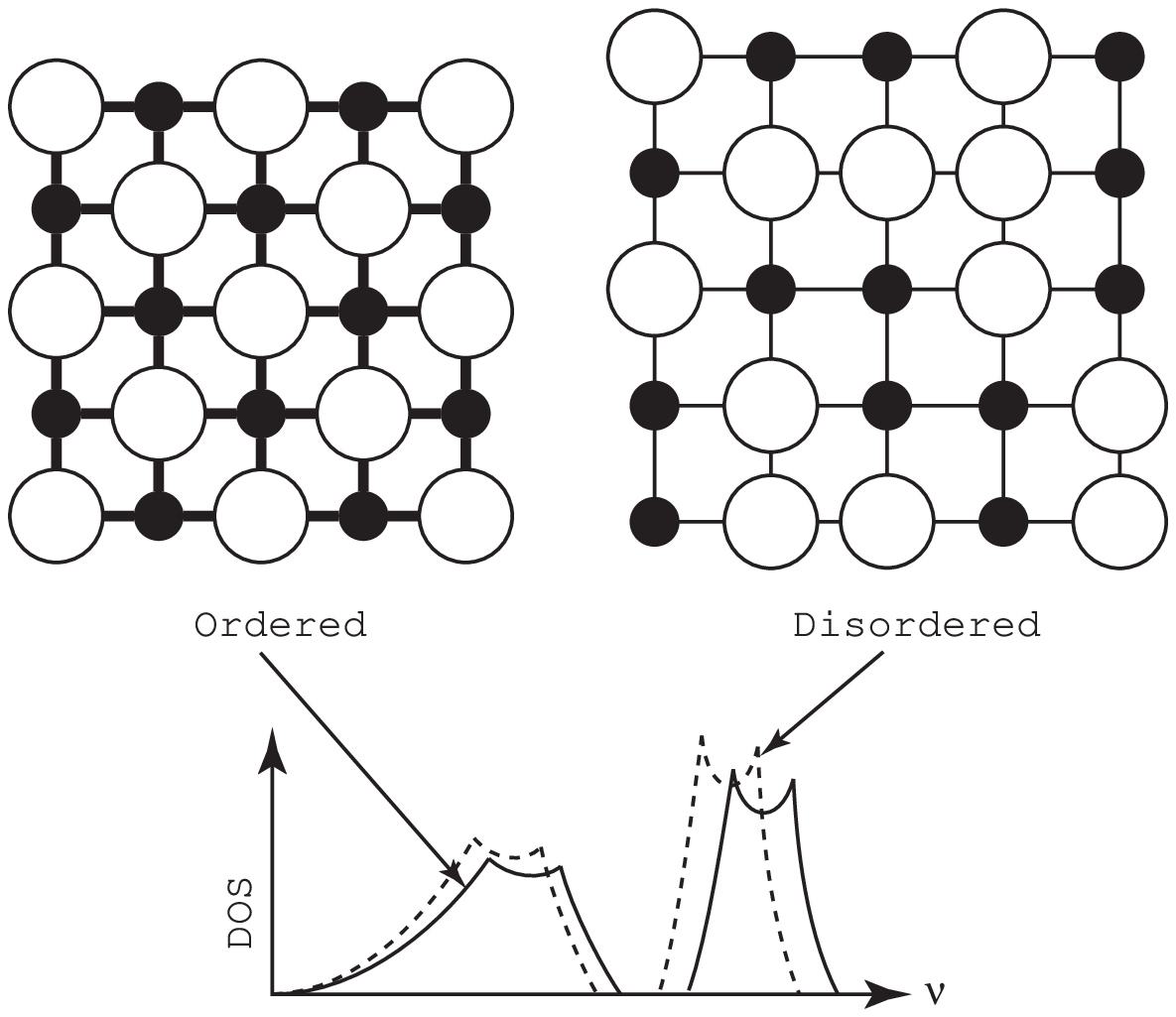}{The Volume Mechanism.
Disordering is typically associated with an increase in volume.
Since chemical bonds tend to soften as they lengthen, an overall
increase in volume should be associated with a corresponding decrease
in the frequency of all phonon modes, resulting in an increase
in vibrational entropy.
}

\illuseps{3.4in}{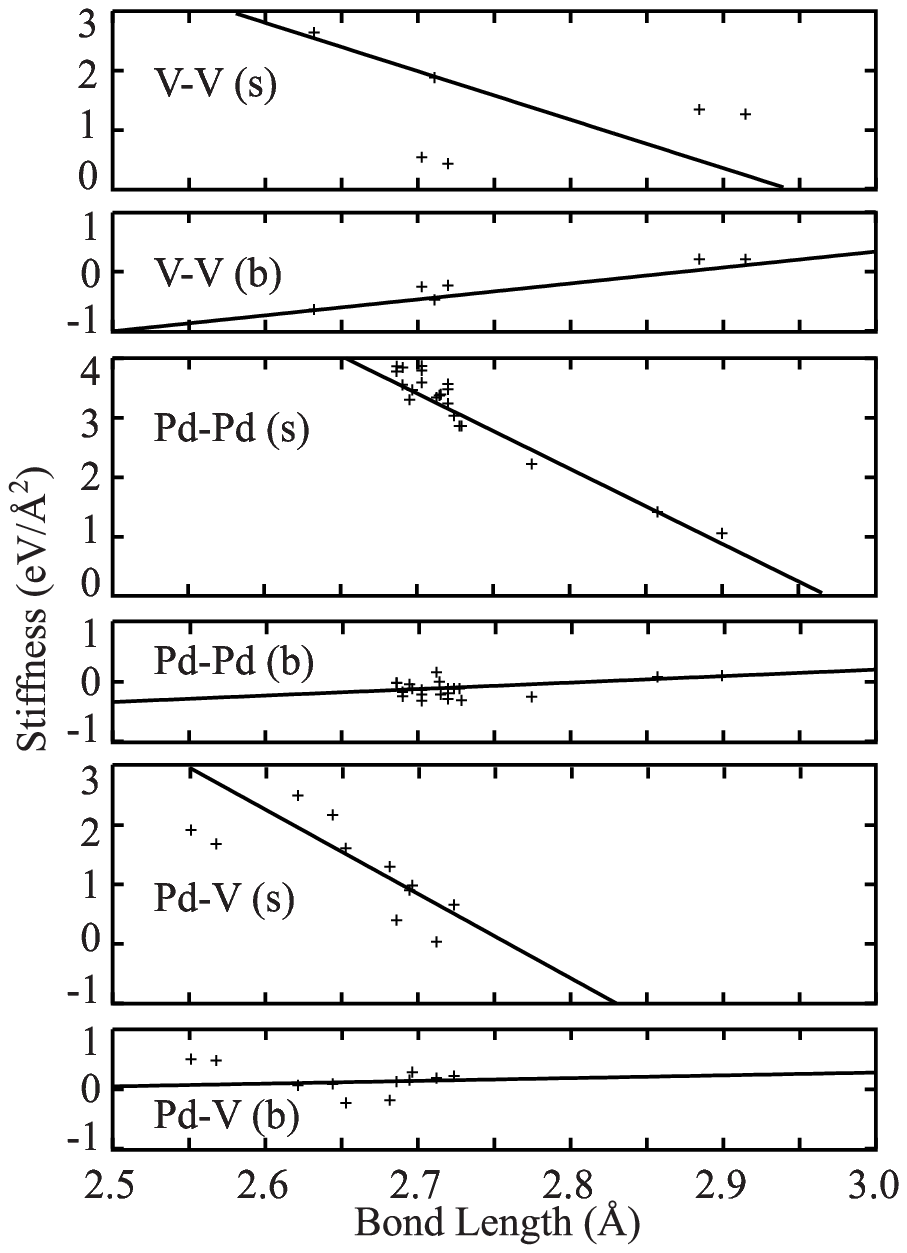}{Stretching
(s) and Bending (b) Terms of the Nearest-Neighbor
Spring Tensor as a Function of Bond Length. Each point corresponds to
one type of bond in one of a set of fcc structures
(L1$_2$, D0$_{22}$, SQS-8, fcc Pd and fcc V,
each taken at two different volumes).}

\illuseps{3.4in}{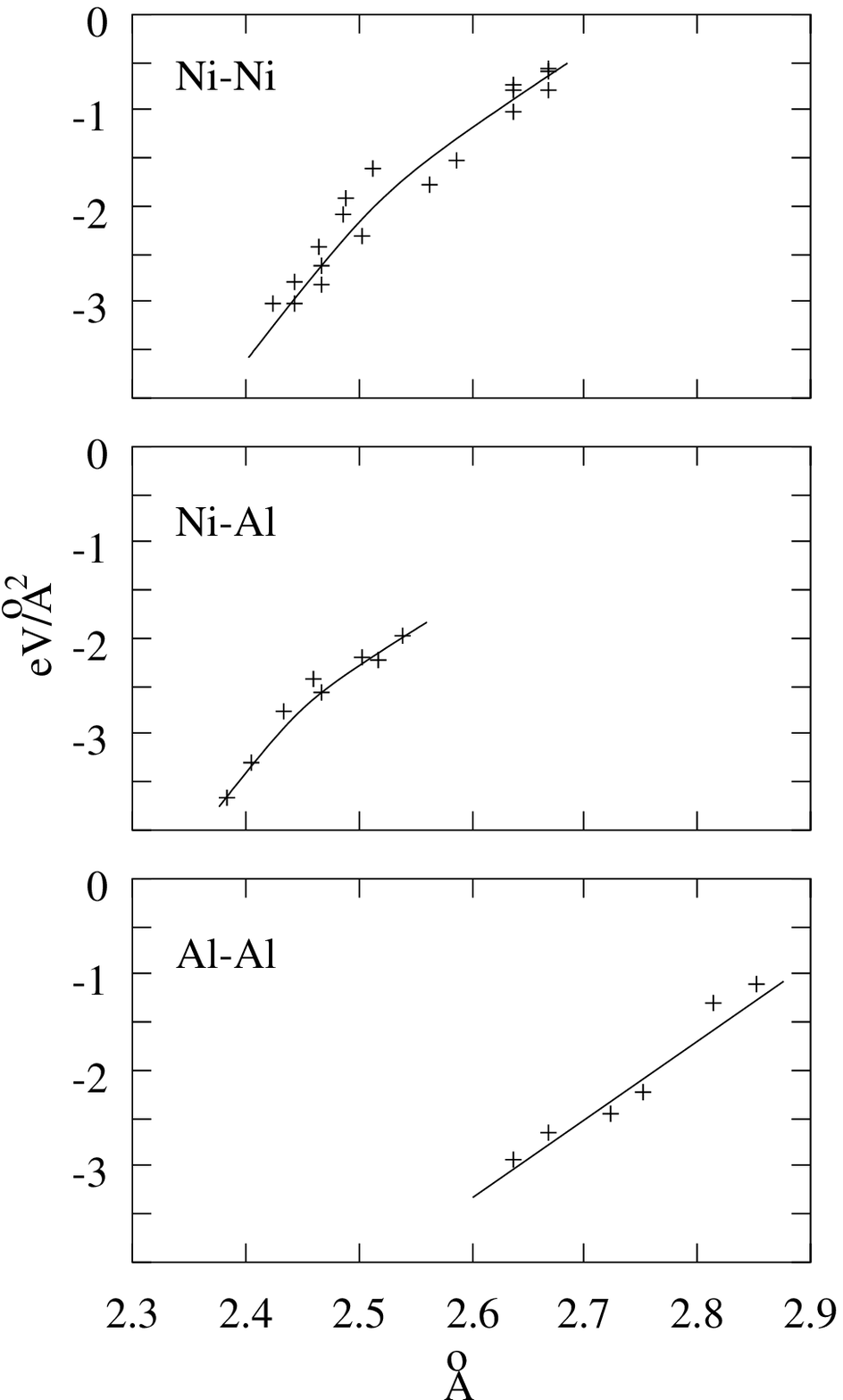}{
Bond Stiffness as a Function of Bond Length in the Ni-Al System.
Only stretching terms are shown. Solid lines are the result of a fit to a
second order polynomial. Each point corresponds to
one type of bond in one of a set of fcc structures
(L1$_2$, D0$_{22}$, SQS-8, fcc Al and fcc Ni,
each taken at two different volumes)}

\illuseps{3.4in}{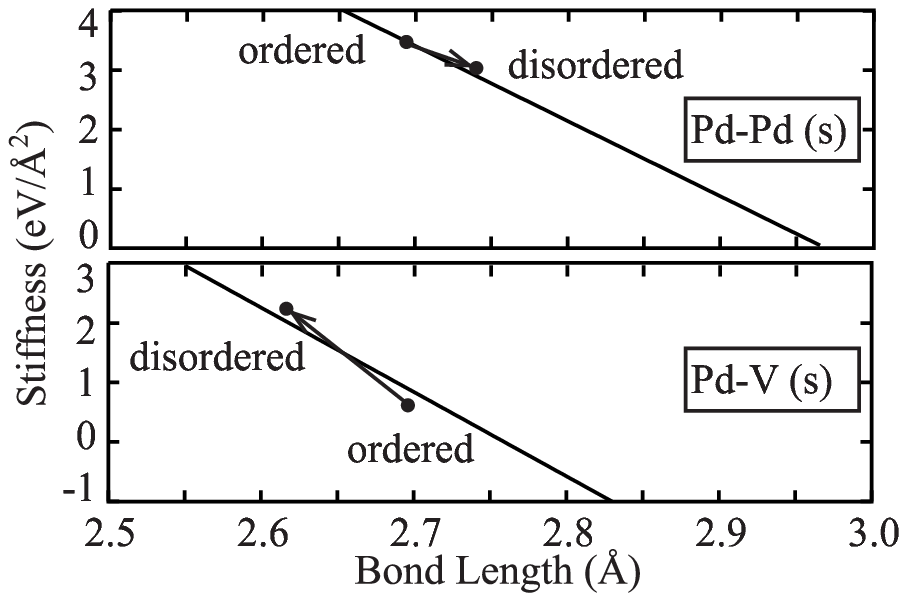}{Shift in Average Bond Stiffness (Along the Stretching Direction) and 
Bond Length upon Disordering.
The fitted line of Fig. \protect\ref{fitpp} is shown for reference.}

\end{document}